\documentclass[%
aps,
 jmp,%
 amsmath,amssymb,
 reprint,%
]{revtex4-1}
\usepackage{graphicx}% Include figure files
\usepackage{subfigure}
\usepackage{mathrsfs}
\usepackage{multirow}
\usepackage[bookmarks=true,colorlinks=true,citecolor=blue,linkcolor=red,urlcolor=magenta]{hyperref}
\usepackage{dcolumn}% Align table columns on decimal point
\usepackage{bm}% bold math
\graphicspath{{figures_pdf_part3/}}

\begin{document}

%\preprint{AIP/123-QED}
\preprint{APS/123-QED}

%\title[Sample title]
\title[]{Mechanisms of proton-proton inelastic cross-section growth in multi-peripheral model within the framework of perturbation theory. Part 3}% Force line breaks with \\
%\thanks{Footnote to title of article.}

\author{I.V. Sharf}
% \altaffiliation[Also at ]
% \altaffiliation{Odessa National Polytechnic University, Shevchenko av. 1, Odessa, 65044, Ukraine.}%Lines break automatically or can be forced with \\
\affiliation{ Odessa National Polytechnic University, Shevchenko av. 1, Odessa, 65044, Ukraine.}%

\author{G.O. Sokhrannyi}%
\affiliation{ Odessa National Polytechnic University, Shevchenko av. 1, Odessa, 65044, Ukraine.}%

\author{A.V. Tykhonov}%
\affiliation{ Odessa National Polytechnic University, Shevchenko av. 1, Odessa, 65044, Ukraine.}%
\affiliation{ Department of Experimental Particle Physics, Jozef Stefan Institute,% \\
     Jamova 39, SI-1000 Ljubljana, Slovenia.}%

\author{K.V. Yatkin}%
\affiliation{ Odessa National Polytechnic University, Shevchenko av. 1, Odessa, 65044, Ukraine.}%

\author{N.A. Podolyan}%
\affiliation{ Odessa National Polytechnic University, Shevchenko av. 1, Odessa, 65044, Ukraine.}%

\author{M.A. Deliyergiyev}%
\affiliation{ Odessa National Polytechnic University, Shevchenko av. 1, Odessa, 65044, Ukraine.}%
\affiliation{ Department of Experimental Particle Physics, Jozef Stefan Institute,% \\
     Jamova 39, SI-1000 Ljubljana, Slovenia.}%

\author{V.D. Rusov}%
 \email{siiis@te.net.ua}
\affiliation{ Odessa National Polytechnic University, Shevchenko av. 1, Odessa, 65044, Ukraine.}%
\affiliation{Department of Mathematics, Bielefeld University, %\\
      Universitatsstrasse 25, 33615 Bielefeld, Germany.}%

% \homepage{http://www.Second.institution.edu/~Charlie.Author.}
%\affiliation{%
%Second institution and/or address%\\This line break forced% with \\
%}%

\date{\today}% It is always \today, today,

\begin{abstract}
We develop a new method for taking into account the interference contributions to proton-proton inelastic cross-section within the framework of the simplest multi-peripheral model based on the self-interacting scalar ${\phi ^3}$ field theory, using Laplace's method for calculation of each interference contribution.

We do not know any works that adopted the interference contributions for inelastic processes. This is due to the generally adopted assumption that the main contribution to the integrals expressing the cross section makes multi-Regge domains with its characteristic strong ordering of secondary particles by rapidity. However, in this work, we find what kind of space domains makes a major contribution to the integral and these space domains are not multi-Regge. We demonstrated that because these interference contributions are significant, so they cannot be limited by a small part of them.  With the help of the approximate replacement the sum of a huge number of these contributions by the integral were calculated partial cross sections for such numbers of secondary particles for which direct calculation would be impossible.

% It was demonstrated in \onlinecite{part1} that accounting all significant interference contributions can lead to virtuality reduction with the energy growth $\sqrt s$, which can results in the experimentally observed increase of total cross-section of hadrons. 
The offered model qualitative agrees with experimental dependence of total scattering cross-section on energy $\sqrt s $ with a characteristic minimum in the range $\sqrt s  \approx 10$ GeV.
%The behavior of the sum of partial cross sections agrees qualitatively with the experimental dependence of the total cross section of hadrons on the energy. 
However, quantitative agreement was not achieved; we assume that due to the fact that we have examined the simplest diagrams of $\phi^3$ theory.
%the considered in \onlinecite{part1} mechanism of virtuality diminishing at the energy $\sqrt s $ growth with consideration of all considerable interference contributions into account can be responsible for the total hadron scattering cross-section growth which is experimentally observed. 
\end{abstract}

%Use showkeys class option if keyword
\maketitle

%\begin{quotation}
%The ``lead paragraph'' is encapsulated with the \LaTeX\
%\verb+quotation+ environment and is formatted as a single paragraph before the first section heading.
%(The \verb+quotation+ environment reverts to its usual meaning after the first sectioning command.)
%Note that numbered references are allowed in the lead paragraph.
%
%The lead paragraph will only be found in an article being prepared for the journal \textit{Chaos}.
%\end{quotation}

%\section{\label{sec:level1}First-level heading:\protect\\ The line
%break was forced \lowercase{via} \textbackslash\textbackslash}

%##################################################################
\section{Introduction}
\label{SECTION_Intro}
%##################################################################

This paper is the sequel to [\onlinecite{part1, part2}], where to calculate proton-proton scattering partial cross-sections within the framework of multi-peripheral model the Laplace method was applied.

\begin{figure}
\begin{center}
  \includegraphics[scale=0.65]{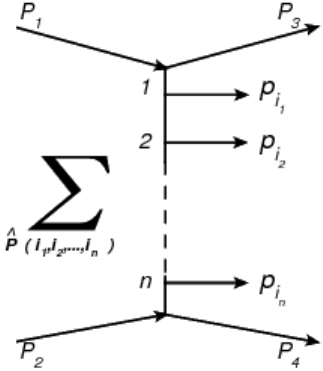} 
\end{center}
 \caption{ Diagram representation of an inelastic scattering amplitude when the $n$ secondary particles are formed. Here $P_1$ and $P_2$ are the four-momenta of primary particles before scattering; $P_3$ and $P_4$ are the four-momenta of primary particles after scattering; ${p_{{i_1}}},{p_{{i_2}}}, \cdots ,{p_{{i_n}}}$ are the four-momenta of secondary particles. Symbol $\sum\limits_{\hat P({i_1},\;{i_2},...,\,{i_n})} {} $ denote a sum over all permutations of indices  ${i_1} = 1,{i_2} = 2,...,{i_n} = n$. }
  \label{fig:part3_fig01} 
\end{figure}

The inelastic scattering amplitude with production of a specified multiplicity of secondary particles, in framework of the multi-peripheral model can be represented as a sum of diagrams demonstrated on Fig.\ref{fig:part3_fig01}.
%%%%%%%%%%%%---------------------------------%%%%%%%%%
%%%%%%%%%%%%---------------------------------%%%%%%%%%
%%%%%%%%%%%%---------------------------------%%%%%%%%%
%%%%%%%%%%%%---------------------------------%%%%%%%%%
To calculate the partial cross-section ${\sigma _n}$ is necessary to evaluate an integral of the squared modulus of a sum of contributions shown in Fig.\ref{fig:part3_fig01}. After simple transformations [\onlinecite{part2}], the expression for the partial cross-section can be represented as a sum of ``cut'' diagrams in Fig.\ref{fig:part3_fig02}. We call summands entering into the sum Fig.\ref{fig:part3_fig02} the interference contributions. Approximate calculation of their sum is the purpose of this paper.

At present time the inelastic scattering processes are considered without the interference contributions [\onlinecite{bfkl_1976, PhysRevD.80.045002}]. This due to the generally adopted assumption that the main contribution to the integrals expressing an inelastic processes makes multi-Regge domains [\onlinecite{bfkl_1976, PhysRevD.80.045002, KozlovNSU_2007, Danilov2006187}] with its characteristic strong ordering of secondary particles by rapidity. This means that the rapidity of neighboring particles on the ``comb'' should be different from each other by a large value. Thus the amplitude of the right-hand and left-hand parts of the diagram on Fig.\ref{fig:part3_fig02} for different orders of connecting lines would be significantly different from zero to almost non-overlapping regions of phase space and integral of their product would be a small quantity.

However, as it was shown in [\onlinecite{part1}] near the threshold of the $n$ particles production at the maximum point of the scattering amplitude Fig.\ref{fig:part3_fig01} 
difference between neighboring particle`s of rapidities is close to zero and at higher energies increases logarithmically with energy $\sqrt s$ growth. This difference has factor $1/(n+1)$, so for high numbers of secondary particles it increases slowly with energy. Moreover, even if each of interference terms is insignificant, all of them are positive and a huge amount $n!$ of them not only makes it impossible to discard them, 
but also leads to the conclusion that the contribution of a ``ladder" diagram Fig.\ref{fig:part3_fig02}, which is usually only taken into account, is negligibly small compared with the sum of the remaining interference terms. This was shown in [\onlinecite{part2}]. For the relatively small number of secondary particles ($n\leq8$) we are able to calculate all the interference contributions in the direct way without any approximations.

Further in this paper we will demonstrate method for approximate calculation of the sum of the interference contributions for large numbers of secondary particles, when direct numerical calculation is not feasible.

%##################################################################
\section{Method description}
\label{SECTION_MethodDecription}
%##################################################################
%%%%%%%%%%%%---------------------------------%%%%%%%%%
%%%%%%%%%%%%---------------------------------%%%%%%%%%
\begin{figure}
\begin{center}
  \includegraphics[scale=0.55]{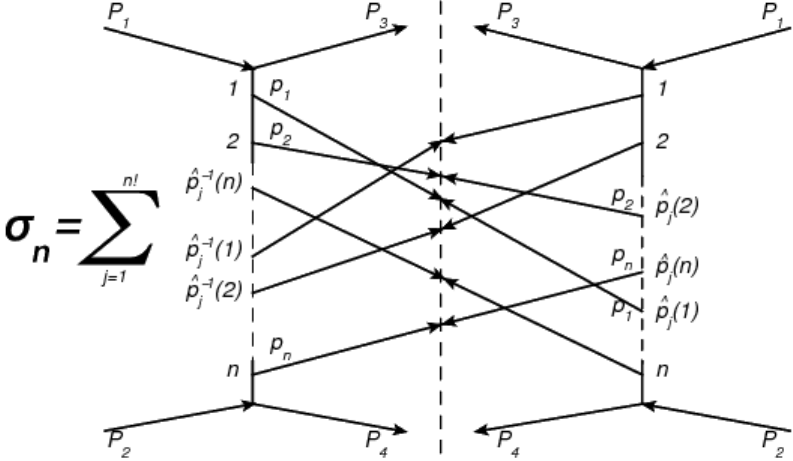} 
\end{center}
 \caption{Representation of the partial cross-section as a sum of ``cut" diagrams. The order of joining of lines with four-momenta $p_k$ from the left-hand side of the cut is as following: the line with $p_1$ is joined to the first vertex, the lines with $p_2$ is joined to the second vertex, etc. The order of joining of lines from the right side of cut corresponds to one of the $n!$ possible permutations of the set of numbers $1, 2,\ldots,n$. Where $\hat{P}_j(k),k=1,2,\ldots,n$ denotes the number into which a number $k$ goes due to permutation $\hat{P}_j$. An integration is performed over the four-momenta $p_k$ for all ``cut lines" taking into account the energy-momentum conservation law and mass shell condition for each of $p_k$. }
  \label{fig:part3_fig02} 
\end{figure}
%%%%%%%%%%%%---------------------------------%%%%%%%%%
%%%%%%%%%%%%---------------------------------%%%%%%%%%

Using the Laplace`s method we have found [\onlinecite{part1, part2}] the mechanism of partial cross-section growth, which was not taken into account in the previously known variants of multi-peripheral model. This mechanism may be responsible for the experimentally observed increase of hadron-hadron total cross-section. However, in this approach based on the Laplace`s method, it was found out that the calculation of partial cross-sections in the multi-peripheral model can be limited just to contributions from the ``cut ladder diagram''. Because for any number of the secondary particles $n$ there is the wide range of energies $\sqrt s $, where such contribution is negligibly small compared to the sum of $n!$ positive interference contributions. At the same time, as we will demonstrated further, the allowance for the interference contributions results in the appearance of multipliers in expression for the partial cross-section, which are decrease with the energy $\sqrt s $ rise (see below Eq.\ref{eq4}). Thereupon the question arises: ``Will the sum of partial cross-sections increase with energy rise if we take interference summands into account?"

As shown in [\onlinecite{part1}], each term in sum shown in Fig.\ref{fig:part3_fig01} with accuracy up to the fixed factor is a function with real and positive values, which has a constrained maximum if its arguments satisfy the mass-shell conditions and energy-momentum conservation law. Therefore, in the c.m.s. of initial particles function corresponding to the left-hand part of cut diagram in Fig.\ref{fig:part3_fig02} can be rewritten in the neighborhood of maximum point in the form [\onlinecite{part1, part2}]
\begin{eqnarray}
&& A\left( {\hat X} \right) = A\left( {{{\hat X}^{\left( 0 \right)}}} \right) \nonumber\\ 
&& \mbox{\fontsize{11}{14}\selectfont $ \times \exp \left( { - \frac{1}{2}{{\left( {\hat X - {{\hat X}^{\left( 0 \right)}}} \right)}^T}\hat D\left( {\hat X - {{\hat X}^{\left( 0 \right)}}} \right)} \right)$ } 
\label{eq1}
\end{eqnarray}%
where $\hat X$ is the column composed of $3n+2$ independent variables, on which the scattering amplitude depends after consideration of mass-shell conditions and energy-momentum conservation law; the first $n$ components of column are the rapidities of secondary particles; the next $n$ components are the $x$ components of transversal momenta of secondary particles (it is supposed that the reference system is chosen so that $Z$-axis is directed in the line of the three-dimensional momentum $P_1$ of initial particle in Fig.\ref{fig:part3_fig01}), the $y$ - components of secondary particle transversal momenta and the two last variables are the antisymmetric combinations of particle transversal momenta $P_3$ and $P_4$, i.e., 
${X_{3n + 1}} = \frac{1}{2}\left( {{P_{3 \bot x}} - {P_{4 \bot x}}} \right)$ 
\\
${X_{3n + 2}} = \frac{1}{2}\left( {{P_{3 \bot y}} - {P_{4 \bot y}}} \right)$
%---------------------------------------------------------------------------------
%\begin{eqnarray}
%&& {X_{3n + 1}} = \frac{1}{2}\left( {{P_{3 \bot x}} - {P_{4 \bot x}}} \right) \nonumber\\ 
%&& {X_{3n + 2}} = \frac{1}{2}\left( {{P_{3 \bot y}} - {P_{4 \bot y}}} \right) 
%\label{eq_2a}
%\end{eqnarray}%
%---------------------------------------------------------------------------------
We denote the column of the values of variables in a maximum point through ${\hat X^{\left( 0 \right)}}$ and a matrix with the elements
%---------------------------------------------------------------------------------
\begin{eqnarray}
&& \mbox{\fontsize{11}{14}\selectfont $  {D_{ab}} =  - {\left. {\frac{{{\partial ^2}}}{{\partial {X_a}\partial {X_b}}}\left( {\ln \left( {A\left( {\hat X} \right)} \right)} \right)} \right|_{\hat X = {{\hat X}^{\left( 0 \right)}}}} $} 
 \label{eq_2b}
 \end{eqnarray}%
 %---------------------------------------------------------------------------------
 where
 %---------------------------------------------------------------------------------
 \begin{eqnarray}
&& \mbox{\fontsize{11}{14}\selectfont $ a = 1,2, \cdots ,3n + 2,b = 1,2, \ldots ,3n + 2$}
 \label{eq_2c}
\end{eqnarray}%
%---------------------------------------------------------------------------------
are the coefficients of the Taylor series expansion of amplitude logarithm in the neighborhood of maximum point. As it was shown in [\onlinecite{part1}], if we do our computations in the c.m.s.of initial particles, the maximum is reached when transversal momenta is zero and secondary particle rapidities are close to numbers that formed an arithmetic progression.

If we denote the difference of this progression through $\Delta y\left( {n, \sqrt s} \right)$ and the value of particle`s rapidity to which the line attached to the $k$-th vertex of diagram in Fig.\ref{fig:part3_fig01} corresponds, through
%---------------------------------------------------------------------------------
\begin{eqnarray}
&& \mbox{\fontsize{10}{14}\selectfont $\Delta y\left( {n,\sqrt s} \right) = y_k^{\left( 0 \right)} - y_{k + 1}^{\left( 0 \right)},\quad k = 1,2, \cdots ,n - 1$ }
\label{eq2a}
\end{eqnarray}%
%---------------------------------------------------------------------------------
we get [\onlinecite{part1}]:
%---------------------------------------------------------------------------------
\begin{eqnarray}
&& \mbox{\fontsize{10}{14}\selectfont $ y_k^{\left( 0 \right)} = \left( {\frac{{n + 1}}{2} - k} \right)\Delta y\left( {n, \sqrt s} \right), \quad k = 1,2, \cdots ,n$}  
\label{eq2}
\end{eqnarray}%
%---------------------------------------------------------------------------------
The form of the function $\Delta y\left( {n, \sqrt s} \right)$ has been discussed in [\onlinecite{part1}]. For further consideration, it is important that it is a slowly increasing function on $s$ and decreasing function on the number $n$ of the secondary particles and vanishes when $s$ is equal to the threshold of $n$ particle production. Thus, the column ${\hat X^{\left( 0 \right)}}$ contains only the first nonzero $n$ rapidity components, which are defined by Eq.\ref{eq2}.

The following expression corresponds to the right-hand part of cut diagram in Fig.\ref{fig:part3_fig02}:
%---------------------------------------------------------------------------------
\begin{eqnarray}
&& \mbox{\fontsize{10}{14}\selectfont $ {{\hat P}_j}\left( {A\left( {\hat X} \right)} \right) = A\left( {{{\hat X}^{\left( 0 \right)}}} \right)$ } \nonumber\\ 
&& \times \mbox{\fontsize{10}{14}\selectfont $ \exp \left( { - \frac{1}{2}{{\left( {{{\hat P}_j}\hat X - {{\hat X}^{\left( 0 \right)}}} \right)}^T}\hat D\left( {{{\hat P}_j}\hat X - {{\hat X}^{\left( 0 \right)}}} \right)} \right)$ } \label{eq3}
\end{eqnarray}%
%---------------------------------------------------------------------------------
The interference contribution corresponding to total ``cut'' diagram, which refers to the $j$-th summand in  Fig.\ref{fig:part3_fig02}, is proportional to an integral of the product of functions Eq.\ref{eq1} and Eq.\ref{eq3} over all variables. Denoting an interference summand corresponding to the permutation ${\hat P_j}$ through ${\sigma '_n}\left( {{{\hat P}_j}} \right)$ and calculating its Gaussian integral (at the same time, 
other multipliers besides the squared modulus of scattering amplitude in an integrand are approximately substituted for their values at the maximum point [\onlinecite{part2}]), we have
%other multipliers besides the squared modulus of scattering amplitude in an integrand are approximately replaced by their values at the maximum point [\onlinecite{part2}]), we get 
%---------------------------------------------------------------------------------
\begin{eqnarray}
&& {\sigma '_n}\left( {{{\hat P}_j}} \right) = \frac{{{{\left( {A\left( {{{\hat X}^{\left( 0 \right)}}} \right)} \right)}^2}v\left( {\sqrt s } \right)}}{{\sqrt {\det \left( {\frac{1}{2}\left( {\hat D + \hat P_j^T\hat D{{\hat P}_j}} \right)} \right)} }} \nonumber\\ 
&& \times \exp \left( { - \frac{1}{2}\left( {{{\left( {\Delta \hat X_j^{\left( 0 \right)}} \right)}^T}{{\hat D}^{\left( j \right)}}\Delta \hat X_j^{\left( 0 \right)}} \right)} \right) 
\label{eq4} 
\end{eqnarray}%
%---------------------------------------------------------------------------------
where we use the following notations:
%--------------------------------------------------------------------------------
\begin{subequations}
\begin{eqnarray}
&& \Delta \hat X_j^{\left( 0 \right)} = {\hat X^0} - {\hat P_j}^{ - 1}\left( {{{\hat X}^{\left( 0 \right)}}} \right) \\
&& {\hat D^{\left( j \right)}} = {\left( {{{\hat D}^{ - 1}} + {{\hat P}_j}^T{{\hat D}^{ - 1}}{{\hat P}_j}} \right)^{ - 1}} \\
&& \mbox{\fontsize{11}{14}\selectfont $ v\left( {\sqrt s } \right) = \frac{1}{2}\frac{1}{{\sqrt s \sqrt {s/4 - {M^2}} \left( {E_P/2} \right)\sqrt {{{\left( {E_P/2} \right)}^2} - {M^2}} }}$ } \\
&& {E_P} = \sqrt s  - \sum\limits_{k = 1}^n {{\mathop{\rm ch}\nolimits} \left( {y_k^{\left( 0 \right)}} \right)} 
\end{eqnarray}
\end{subequations}
%---------------------------------------------------------------------------------

$M$ is the mass of initial particle, which is made dimensionless by the mass of secondary particle (it is supposed that the energy $\sqrt s $ is also made dimensionless by the mass of the secondary particle). %The magnitude designated as $const$ contains fixed factors insignificant for our consideration \onlinecite{part2}. Therefore, we examine the interference contributions nondimensionalized by this const. 

Note, that here and in the following sections we will use the ``prime" sign in ours notation to indicate that we use
a dimensionless quantity that characterized the dependence of the cross-sections on energy, but not their absolute values.

The value of amplitude at the maximum point $A\left( {{{\hat X}^{\left( 0 \right)}}} \right)$ increases with the $\sqrt s $ growth due to mechanism of virtuality reduction [\onlinecite{part1}]. However, the distance $\Delta \hat X_j^{\left( 0 \right)}$ between maximum points of ``cut'' diagram also increases with the $\sqrt s $ growth. Therefore, the exponential factor entering in Eq.\ref{eq4} can decrease with energy growth. This makes considered above question. How competition of these two multipliers will result on the dependence of the sum of partial cross-sections on $\sqrt s $?

Thus, each interference contribution can be computed numerically. However due to the huge number of contributions and large number of secondary particles $n$ the direct numerical calculation of the sum of interference terms in Fig.\ref{fig:part3_fig02} is impossible. 
%%%%%%%%%%%%---------------------------------%%%%%%%%%
%%%%%%%%%%%%---------------------------------%%%%%%%%%
\begin{figure}
\begin{center}
  \centering
  \subfigure[]{
  \includegraphics[scale=0.39]{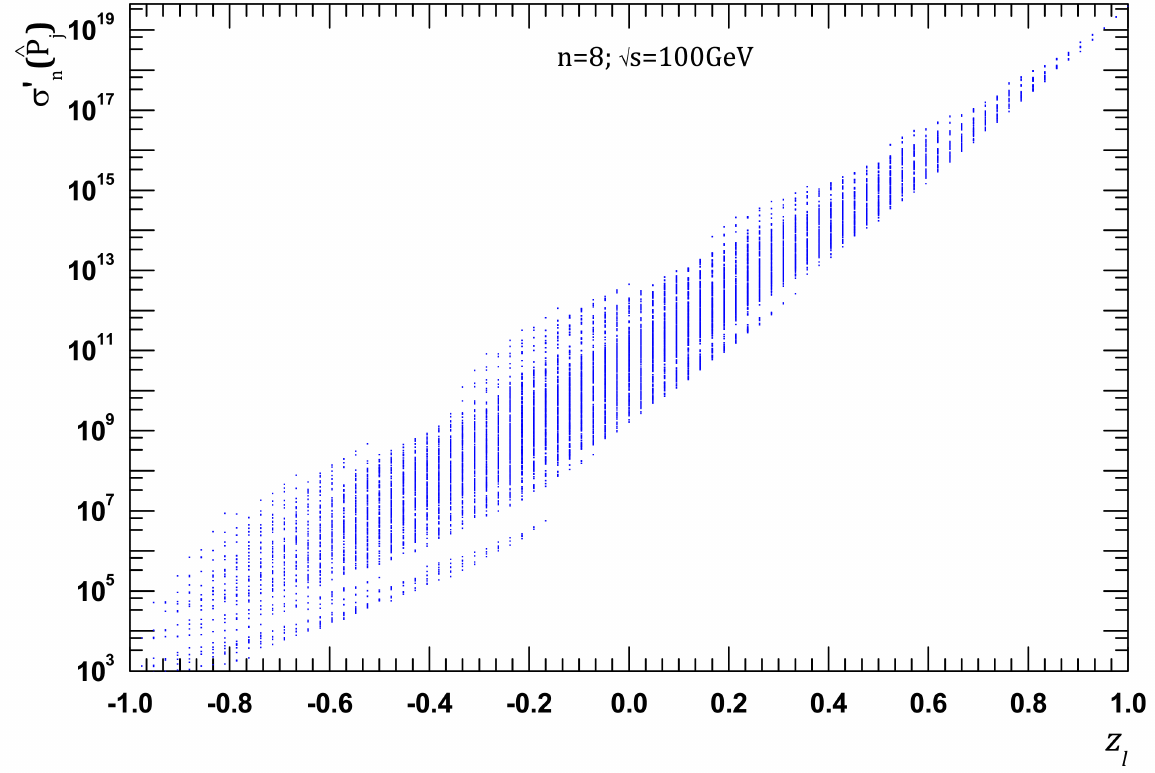} 
  \label{fig:part3_fig03a} 
  }
  \subfigure[]{
  \includegraphics[scale=0.39]{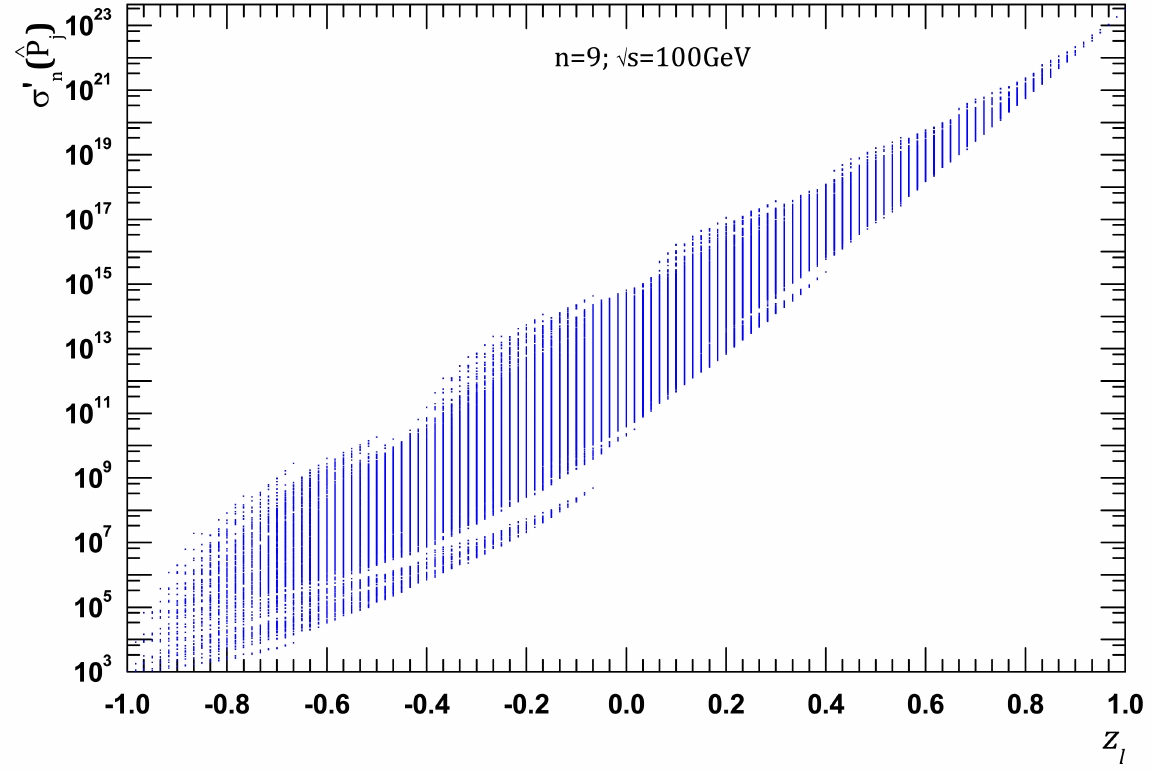}   
  \label{fig:part3_fig03b} 
  }
\end{center}
 \caption{
The magnitude of interference contributions as function of ${z_l}$ at $\sqrt s =1000$ GeV: (a) $n=8$, (b) $n=9$. Here and in further figures the interference contributions indicated on the $y$-axis are divided by the common multiplier $\exp \left( { - \sum\limits_{a = 1}^{3n + 2} {\sum\limits_{b = 1}^{3n + 2} {X_a^{\left( 0 \right)}{D_{ab}}X_b^{\left( 0 \right)}} } } \right)$. 
Obviously, that one value of $z_l$ maps to many different contributions. Moreover,
one can see that the mean values ​​of the logarithms of these contributions are placed approximately along a straight line (see below Eq.\ref{eq16} and Fig.\ref{fig:part3_fig04}).}
  \label{fig:part3_fig03} 
\end{figure}
%%%%%%%%%%%%---------------------------------%%%%%%%%%
%%%%%%%%%%%%---------------------------------%%%%%%%%%
We can avoid this difficulty in the following way. The maximum in the right part of cut diagram in Fig.\ref{fig:part3_fig02} is attained at $\hat{X}=\hat{P}_j^{-1}\left( \hat{X}^{(0)} \right)$. In other words, a maximum of function, which is associated with the right-hand part of cut diagram, can be obtained from a maximum of function, which maps with the left-hand part of cut diagram, by the rearrangement of arguments. Then the value of each interference contribution is determined by the distance between points of maximum in the right-hand and left-hand part of cut diagram as well as by the relative position of these maximum points, since in different directions contributions to scattering amplitude fall off with distance from point of maximum, in general, with different rate, and also by the relative position of proper directions of the matrices $\hat{D}$ and $\hat{P}_j^T\hat{D}\hat{P}_j$. In other words, multiplying Gaussian functions corresponding to the right-hand and to the left-hand part of interference diagrams in Fig.\ref{fig:part3_fig02} each time we will obtain as a result Gaussian function, which has the proper value at the maximum point (which we call the ``height" of the maximum) and the proper multidimensional volume cutout by resulting Gaussian function from an integration domain (which we call the ``width" of the maximum).
%%%%%%%%%%%%---------------------------------%%%%%%%%%
%%%%%%%%%%%%---------------------------------%%%%%%%%%
\begin{figure*}
\begin{center}
  \centering
  \subfigure[]{
  \includegraphics[scale=0.75]{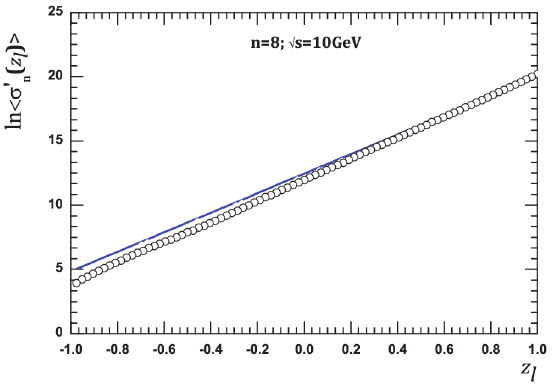} 
  \label{fig:part3_fig04a} 
  }
  \subfigure[]{
  \includegraphics[scale=0.75]{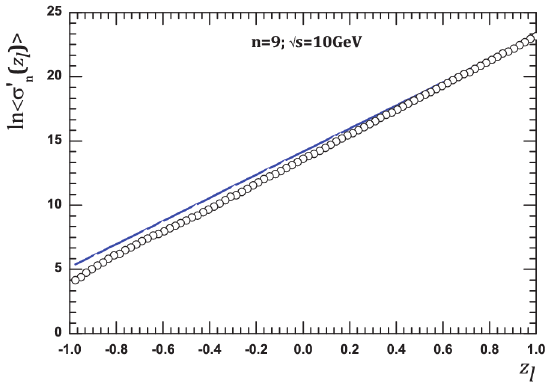}   
  \label{fig:part3_fig04b} 
  }
  \subfigure[]{
  \includegraphics[scale=0.75]{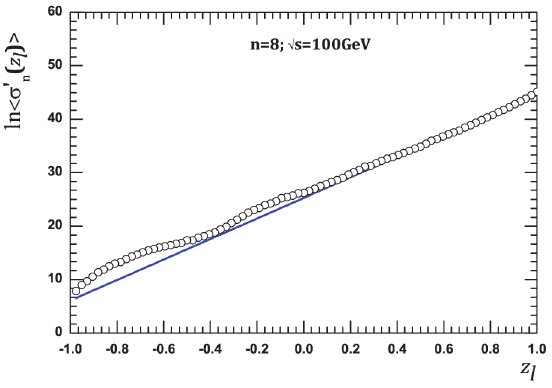} 
  \label{fig:part3_fig04c} 
  } 
  \subfigure[]{
  \includegraphics[scale=0.75]{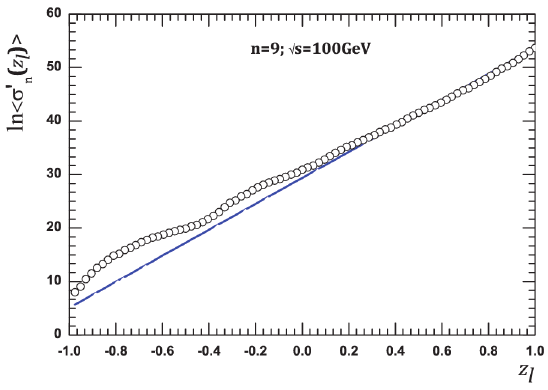} 
  \label{fig:part3_fig04d} 
  } 
\end{center}
 \caption{
Comparing the values of
%Two results of 
$\ln \left( {\left\langle {{\sigma '_n}\left( {{z_l}} \right)} \right\rangle } \right)$ 
obtained by a direct numerical calculation with examination of all interference contributions (circles), and the ones obtained by the approximation Eq.\ref{eq15} (straight line) at $n=8$, $\sqrt s  = 10$ GeV (a); $n=9$, $\sqrt s  = 10$ GeV (b); $n=8$, $\sqrt s  = 100$ GeV (c); $n=9$, $\sqrt s  = 100$ GeV (d).}
  \label{fig:part3_fig04} 
\end{figure*}
%%%%%%%%%%%%---------------------------------%%%%%%%%%
%%%%%%%%%%%%---------------------------------%%%%%%%%%
%%%%%%%%%%%%---------------------------------%%%%%%%%%
%%%%%%%%%%%%---------------------------------%%%%%%%%%
\begin{figure*}
\begin{center}
  \centering
  \subfigure[]{
  \includegraphics[scale=0.35]{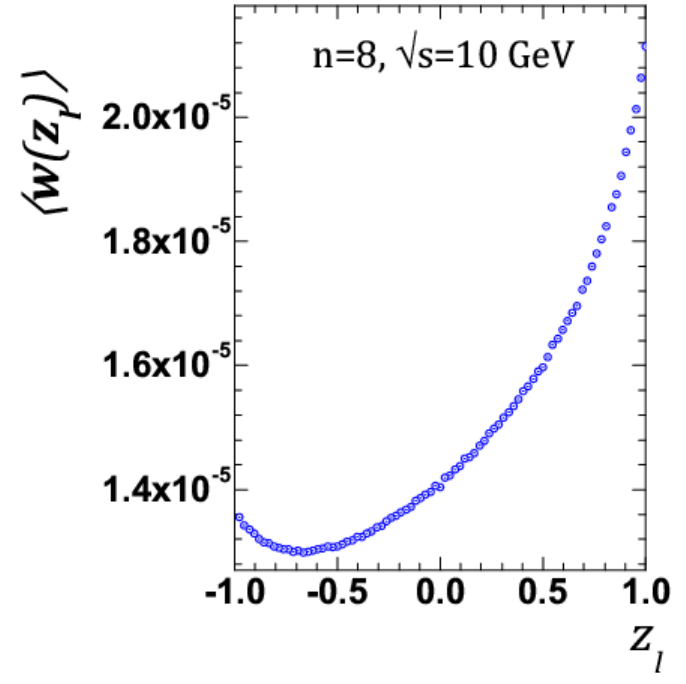} 
  \label{fig:part3_fig05a} 
  }
  \subfigure[]{
  \includegraphics[scale=0.35]{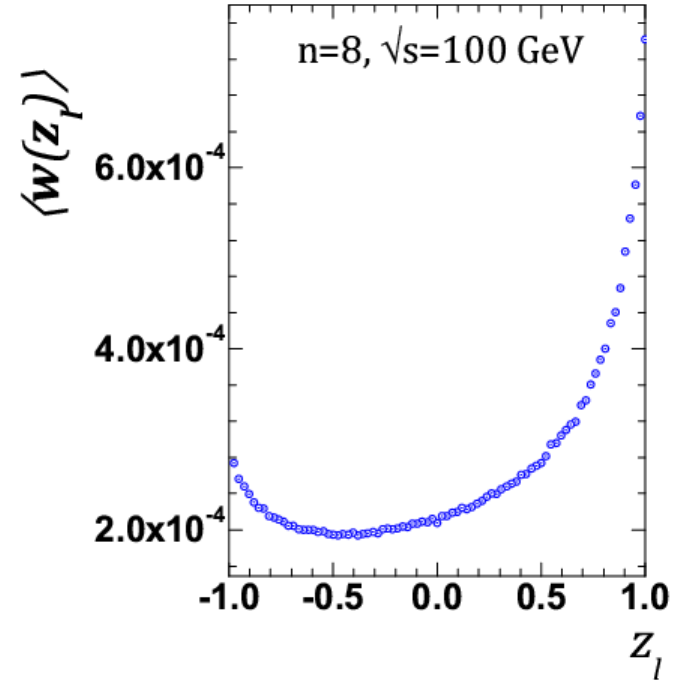} 
  \label{fig:part3_fig05c} 
  } 
  \subfigure[]{
  \includegraphics[scale=0.35]{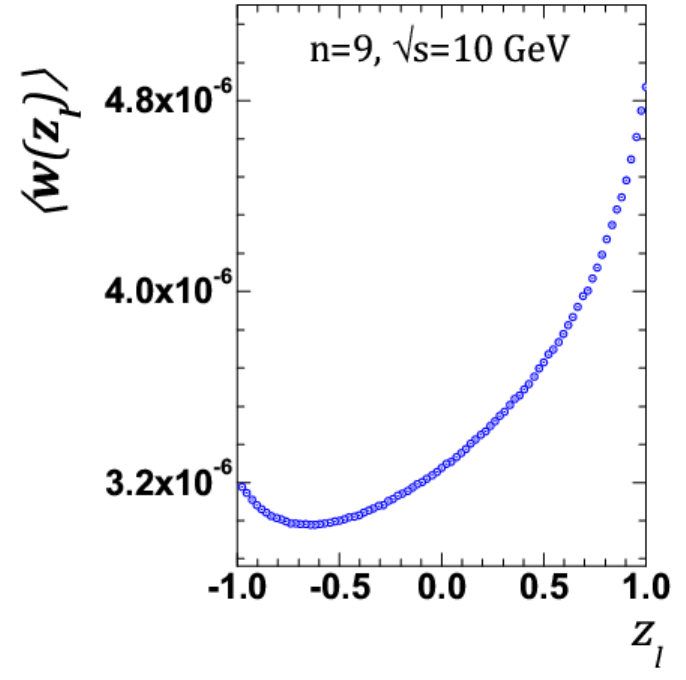} 
  \label{fig:part3_fig05e} 
  }
  \subfigure[]{
  \includegraphics[scale=0.35]{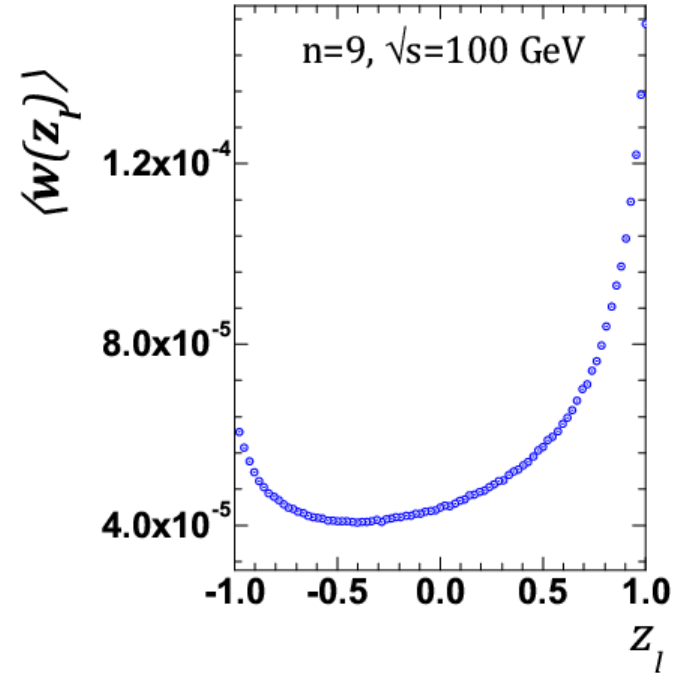} 
  \label{fig:part3_fig05g} 
  } 
  \subfigure[]{
  \includegraphics[scale=0.35]{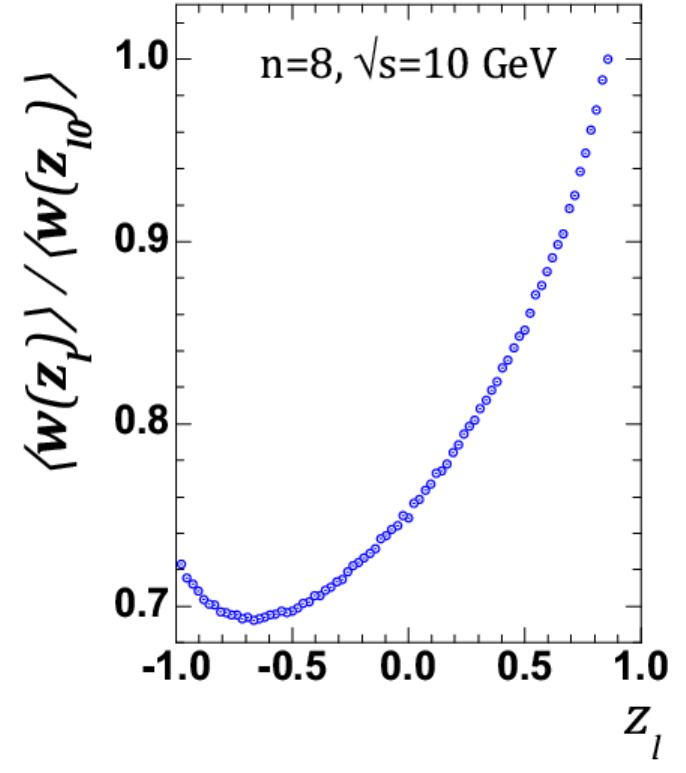}   
  \label{fig:part3_fig05b} 
  }
  \subfigure[]{
  \includegraphics[scale=0.35]{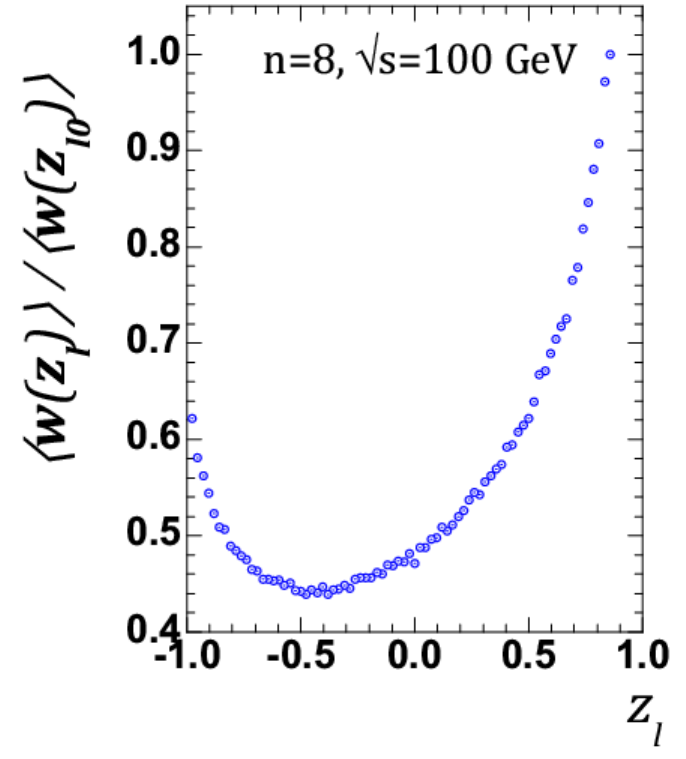} 
  \label{fig:part3_fig05d} 
  } 
  \subfigure[]{
  \includegraphics[scale=0.35]{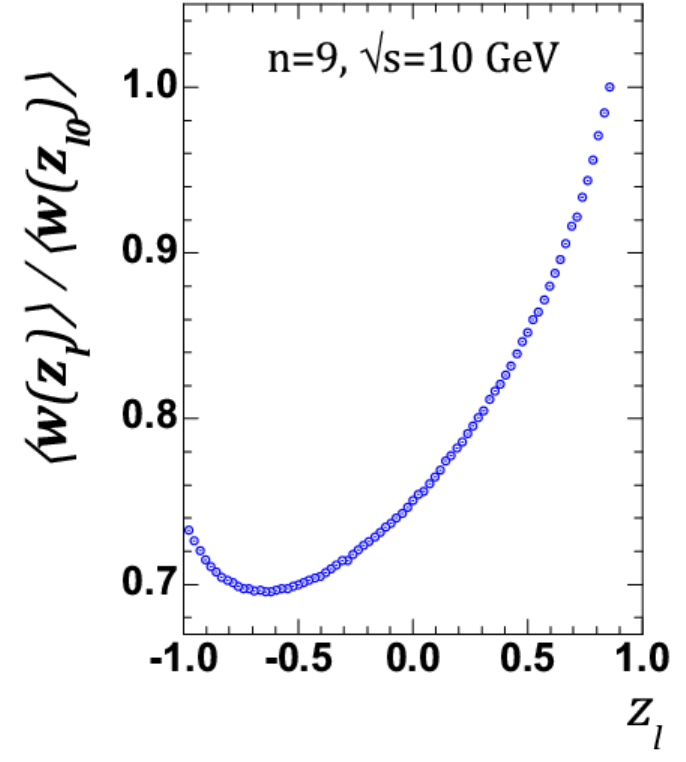}   
  \label{fig:part3_fig05f} 
  }
  \subfigure[]{
  \includegraphics[scale=0.35]{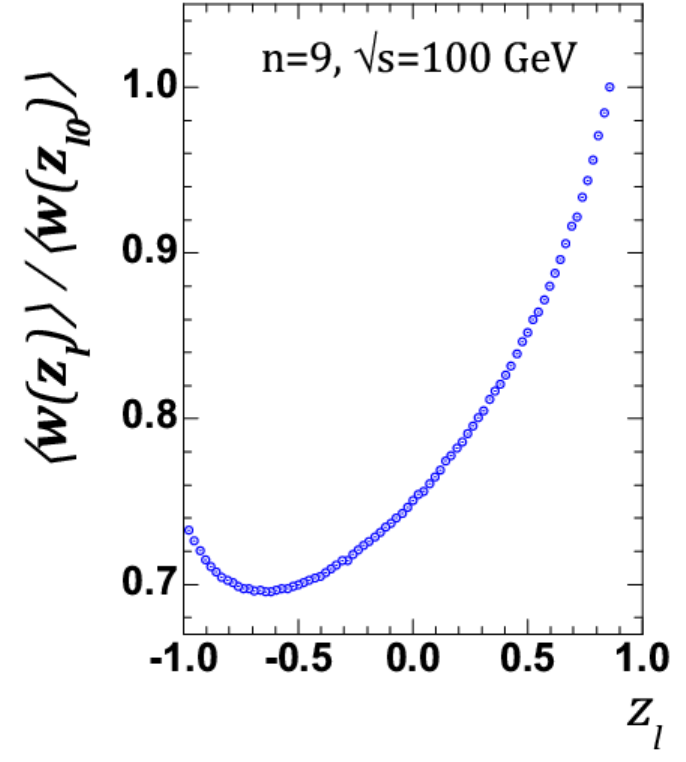} 
  \label{fig:part3_fig05h} 
  } 
\end{center}
 \caption{
The values of $\left\langle {w\left( {{z_l}} \right)} \right\rangle $  obtained by direct calculation values Eq.\ref{eq15} for all interference contributions for $n=8$ and $n=9$ at $\sqrt s  = 10$ GeV \ref{fig:part3_fig05a},\ref{fig:part3_fig05e} accordingly; for the same number $n$, but at $\sqrt s  = 100$ GeV \ref{fig:part3_fig05c},\ref{fig:part3_fig05g} and the ratio $\left\langle {w({z_l}} \right\rangle /\left\langle {w({z_l}_0} \right\rangle $ for $n=8$ and $n=9$ at $\sqrt s  = 10$ GeV \ref{fig:part3_fig05b},\ref{fig:part3_fig05f}; $\sqrt s  = 100$ GeV \ref{fig:part3_fig05d},\ref{fig:part3_fig05h}.}
  \label{fig:part3_fig05} 
\end{figure*}
%%%%%%%%%%%%---------------------------------%%%%%%%%%
%%%%%%%%%%%%---------------------------------%%%%%%%%%

We assume that summands in Fig.\ref{fig:part3_fig02} are arranged in ascending order of the distance between the maximum points in the right-hand part and left-hand part of cut diagram (we denote this distance through $r$) so that "cut" diagram with the initial attachment of lines to the right-hand part of diagram corresponds to $j=1$. In other words, the line of secondary particle with the four-momentum $p_i$ is attached to the $i$-th top in the right-hand part of cut diagram in Fig.\ref{fig:part3_fig02}. As follows from Eq.\ref{eq4}, the interference contributions exponentially decrease with the $r^2$ growth. However, in spite of this the interference contributions do not become negligible due to their huge number, which, as discussed below, are increases very rapidly with $r^2$ growth. The value of $r^2$ is proportional to the square of magnitude $\Delta y(n, \sqrt s)$, which, as was noted above, is zero on the threshold of $n$ particle production and slowly increases with distance from this threshold. Therefore, for each number $n$ there is the fairly wide range of energies close to the threshold, in which the sharpness of decrease of the interference contributions with the $r^2$ increase is small in the sense that it is less important factor than the increase in their number. At such energies, which we call ``low", the partial cross-section $\sigma '_n$ is determined by the sum of huge number of small interference contributions. When the magnitude $\Delta y(n, \sqrt s)$ is increased with the further growth of energy $\sqrt s$, the decrease rate of interference contributions increases, while the growth rate of their number with the $r^2$ increase does not change with energy. At such energies, which we call ``high", the main contribution to the partial cross-section is made by the relatively small number of interference terms corresponding to the small $r^2$, which can be calculated by Eq.\ref{eq4}.

If we compose the $n$-dimensional vector (we denote it through $\vec{y}^{(0)}$) from the particle rapidities Eq.\ref{eq2}, which constrainedly maximizes the function associated with the diagram with the initial arrangement of momenta in Fig.\ref{fig:part3_fig02}, vectors maximizing the functions with another momentum arrangement will differ from the initial vector only by the permutation of components, i.e., these vectors have the same length. Consider two such $n$-dimensional vectors, one of which corresponds to the initial arrangement, and another - to some permutation, then in the $n$-dimensional space it is possible to ``pull on'' a two-dimensional plane on them (as a set of their various linear combinations), where two-dimensional geometry takes place. Therefore, the distance $r$ will be determined by cosine of an angle between the considered equal on length $n$-dimensional rapidity vectors in the two-dimensional plane, ``pulled'' on them. An angle corresponding to the $\hat{P}_j$ permutation we designate through $\theta _j$, $0\leq\theta _j \leq \pi$.

Thus, each of the terms in the sum Fig.\ref{fig:part3_fig02} can be uniquely matched to its angle $\theta _j$. At the same time the variable $z=cos(\theta)$ is more handy for consideration than an angle $\theta _j$. Using Eq.\ref{eq2}, can be shown that the variable $z$ can take discrete set of values:
%---------------------------------------------------------------------------------
\begin{eqnarray}%
&& {z_l} = 1 - \frac{{12}}{{\left( {n - 1} \right)n\left( {n + 1} \right)}}l\quad  \cr 
&& l = 0,1, \cdots ,\frac{{\left( {n - 1} \right)n\left( {n + 1} \right)}}{6} 
\label{eq5} 
\end{eqnarray}%
%---------------------------------------------------------------------------------

Note that although the relation Eq.\ref{eq2} for the rapidities of secondary particles is satisfied with high accuracy at the maximum point, it is still approximate. This means that the contributions, which were matched to the same value as $z$ in Eq.\ref{eq2}, are now matched to values which vary slightly from the values of $z$.

Consequently, the distances between the maximum points in the cut diagrams, which correspond to such contributions, are similar but not equal to each other. In addition, this distance, is not a unique factor which affects the value of interference contribution. Therefore, if each interference contribution is associated to some value of variable $z$ by the approximation Eq.\ref{eq2}
then that the different values of interference contributions correspond to the same value of $z_l$ (see Fig.\ref{fig:part3_fig03}). 

%Consequently, to such contributions correspond a similar but unequal to each other distances between maximum points in a ``cut" diagram. In addition, this distance, as was discussed above, is not a unique factor affecting to the value of interference contribution. Therefore, if each interference contribution is associated with the value of variable $z$ by the approximation Eq.\ref{eq2}, it appears, that the different values of interference contributions correspond to the one and the same value of $z_l$ (see Fig.\ref{fig:part3_fig03}). 

%Thus, while each contribution is associated to some value of variable $z$ in the approximation Eq.\ref{eq2}, the value of contribution is not the unique function of $z$. However, the sum expressing the partial cross-section $\sigma'_n$ can be written in the following way
While each contribution is associated to some value of variable $z$ in the approximation Eq.\ref{eq2}, the value of contribution is not the unique function of $z$. 
The sum which expresses the partial cross-section $\sigma'_n$ can be written as:%---------------------------------------------------------------------------------
\begin{eqnarray}%
&& {\sigma' _n} = \sum\limits_{l = 0}^{\frac{{\left( {n - 1} \right)n\left( {n + 1} \right)}}{6}} {\Delta {N_l}\left( {\frac{{\sum\limits_{{z_j} = {z_l}} {{\sigma '_n}\left( {{{\hat P}_j}} \right)} }}{{\Delta {N_l}}}} \right)} 
\label{eq6}
\end{eqnarray}%
%---------------------------------------------------------------------------------
where $\Delta {N_l}$ is the number of summands to which the value ${z_j} = {z_l}$ is corresponds in the approximation Eq.\ref{eq2}. The mean value of interference contribution in Eq.\ref{eq6} is already the unique function of $z_l$. Therefore, we introduce the following notation
%---------------------------------------------------------------------------------
\begin{eqnarray}%
&& \frac{{\sum\limits_{{z_j} = {z_l}} {{\sigma '_n}\left( {{{\hat P}_j}} \right)} }}{{\Delta {N_l}}} = \left\langle {{\sigma' _n}\left( {{z_l}} \right)} \right\rangle 
\label{eq7}
\end{eqnarray}%
%---------------------------------------------------------------------------------
where $\left\langle \sigma '_n(z_l) \right\rangle$ is some function whose form at ``low" energies can be determined from the following concerns.%considerations.

For any given multiplicity $n$ when the values of parameter $l$ in Eq.\ref{eq5} are small, and when the number of corresponding interference contributions is relatively small, we can directly calculate these elements and their sum. Let`s denote the maximum value of $l$, for which all interference contributions are calculated directly, by $l_0$. In particular, in this paper we managed to calculate the interference contributions up to $l_0=6$. The partial cross-section can be written as: 
%---------------------------------------------------------------------------------
\begin{eqnarray}%
&& {\sigma' _n} = \sigma _n^{\prime(h)} + \sigma _n^{\prime(l)} = \nonumber\\ 
&& \mbox{\fontsize{10}{14}\selectfont $  = \sum\limits_{\scriptstyle {z_j} = {z_l}, \hfill \atop 
  \scriptstyle l = 0,1, \cdots {l_0} \hfill} {{\sigma '_n}\left( {{{\hat P}_j}} \right)} + \sum\limits_{l = {l_0} + 1}^{\frac{{\left( {n - 1} \right)n\left( {n + 1} \right)}}{6}} {\Delta {N_l}\left\langle {{\sigma '_n}\left( {{z_l}} \right)} \right\rangle } $} 
\label{eq8}
\end{eqnarray}%
%---------------------------------------------------------------------------------
where $\sigma_n^{\prime(h)}$ is the sum of contributions sufficient at ``high" energies, and $\sigma_n^{\prime(l)}$ is the sum of contributions sufficient at ``low" energies. Thus, the difficulties in the calculations of the huge number of interference contributions
mainly relates to the range of ``low" energies and can be reduced to the approximate calculation of $\left\langle \sigma '_n(z_l) \right\rangle$ and $\Delta N_l$.
%As one can see from Eq.\ref{eq:eq_part3_4}, the exponential factor exerts the most strong influence on the dependence of $\left\langle \sigma '_n(z_l) \right\rangle$ on $z_l$. Note that the expression $\left( \Delta \hat{X}_j^{(0)}\right)^T \hat{D}^{(j)} \Delta \hat{X}_j^{(0)}$ entering into the exponent in Eq.\ref{eq:eq_part3_4} depends only on those matrix $\hat{D}^{(j)}$ components, which are at the intersection of the first $n$ rows and first $n$ columns, since all column $\Delta \hat{X}_j^{(0)}$ components starting with $n+1$ are zero, because they are the particle momentum transverse components at the maximum point. If we designate a matrix composed of elements located at the intersection of the first $n$ rows and first $n$ columns of the matrix $\hat{D}^{(j)}$ through $\hat{D}_y^{(j)}$ and a matrix, which is obtained from the matrix $\hat{D}$ in analogy, through $\hat{D}_y$, we have 

%##################################################################
\section{The approximate calculation of $\left\langle {{\sigma' _n}\left( {{z_l}} \right)} \right\rangle $.}
\label{SECTION_3}
%##################################################################

As follows from Eq.\ref{eq4}, the exponential factor exerts the most significant effect on the dependence of $\left\langle \sigma '_n(z_l) \right\rangle$ on $z_l$. Note that the expression $\left( \Delta \hat{X}_j^{(0)}\right)^T \hat{D}^{(j)} \Delta \hat{X}_j^{(0)}$ entering into the exponent in Eq.\ref{eq4} depends only on those matrix $\hat{D}^{(j)}$ components, which are at the intersection of the first $n$ rows and first $n$ columns, since all column $\Delta \hat{X}_j^{(0)}$ components starting with $n+1$ are zero, because they are the particle momentum transverse components at the maximum point. If we denote the matrix composed of elements located at the intersection of the first $n$ rows and first $n$ columns of the matrix $\hat{D}^{(j)}$ through $\hat{D}_y^{(j)}$ and a matrix, which is obtained from the matrix $\hat{D}$ in analogy, through $\hat{D}_y$, we have 
%---------------------------------------------------------------------------------
\begin{eqnarray}%
&& {\hat D^{\left( j \right)}}:\hat D_y^{\left( j \right)} = {\left( {\hat D_y^{ - 1} + {{\hat P}_j}^T\hat D_y^{ - 1}{{\hat P}_j}} \right)^{ - 1}}
\label{eq9a}
\end{eqnarray}%
%---------------------------------------------------------------------------------
The matrices $\hat D_y^{ - 1}$ and ${\hat P_j}^T\hat D_y^{ - 1}{\hat P_j}$ have one and the same eigenvalues, but they correspond to different eigenvectors. We denote the normalized to unit eigenvector corresponding to the minimal eigenvalue of matrix $\hat D_y^{ - 1} + {\hat P_j}^T\hat D_y^{ - 1}{\hat P_j}$ through ${\hat u_{\min }}$ and the eigenvalue itself - through ${\lambda _{\min }}$. This implies
%---------------------------------------------------------------------------------
\begin{eqnarray}%
&& \mbox{\fontsize{10}{14}\selectfont $ {\lambda _{\min }} = \hat u_{\min }^T\hat D_y^{ - 1}{{\hat u}_{\min }} + \hat u_{\min }^T{{\hat P}_j}^T\hat D_y^{ - 1}{{\hat P}_j}{{\hat u}_{\min }}$} 
\label{eq9}
\end{eqnarray}%
%---------------------------------------------------------------------------------
Since the minimum eigenvalue of matrix $\hat D_y^{ - 1}$ is equal to the minimum values of quadratic form ${\hat u^T}\hat D_y^{ - 1}\hat u$ for the unit vectors $\hat u$, the magnitude $\hat u_{\min }^T\hat D_y^{ - 1}{\hat u_{\min }}$ is not less than the minimum eigenvalue of matrix $\hat D_y^{ - 1}$. By analogy the magnitude $\hat u_{\min }^T{\hat P_j}^T\hat D_y^{ - 1}{\hat P_j}{\hat u_{\min }}$ is not less than the minimum eigenvalue of matrix ${\hat P_j}^T\hat D_y^{ - 1}{\hat P_j}$, which coincides with the minimal eigenvalue of matrix $\hat D_y^{ - 1}$ and is reciprocal of the maximum eigenvalue of matrix ${\hat D_y}$ denoted through $d_y^{\max }$. Thus, ${\lambda _{\min }} \ge \frac{2}{{d_y^{\max }}}$. From this it follows that, the maximum eigenvalue of matrix ${\left( {\hat D_y^{ - 1} + {{\hat P}_j}^T\hat D_y^{ - 1}{{\hat P}_j}} \right)^{ - 1}}$ does not exceed $d_y^{\max }/2$. By analogy we obtain that the minimum eigenvalue of matrix ${\left( {\hat D_y^{ - 1} + {{\hat P}_j}^T\hat D_y^{ - 1}{{\hat P}_j}} \right)^{ - 1}}$ is no smaller than $d_y^{\min }/2$, where $d_y^{\min }$ is the minimum eigenvalue of matrix ${\hat D_y}$. Thus, an interval enclosing the eigenvalues of matrix $\hat D_y^{\left( j \right)}$ is, at least, twice smaller than an interval enclosing the eigenvalues of matrix ${\hat D_y}$.
We can demonstrate that at approximation of an equal denominators [\onlinecite{part1}] the value of $d_y^{\max }$ can be estimated in the following way
%---------------------------------------------------------------------------------
\begin{eqnarray}%
&& d_y^{\max } \approx \frac{2}{{4{{{\mathop{\rm sh}\nolimits} }^2}\left( {\frac{{\Delta y\left( {n,s} \right)}}{2}} \right) + 1}}
\end{eqnarray}%
%---------------------------------------------------------------------------------
i.e., an interval enclosing the eigenvalues of matrix $\hat D_y^{\left( j \right)}$ at any energies and number of particles is less than unity, whereas at the considerable values of $\Delta y\left( {n,s} \right)$, i.e. at a distance from the threshold, this interval is much less than unity. 

Therefore, if we reduce matrix $\hat D_y^{\left( j \right)}$ to diagonal form, it will be close to a matrix multiple of unit matrix. If we represent this matrix in the form 
%---------------------------------------------------------------------------------
\begin{eqnarray}%
&& \hat D_y^{\left( j \right)} = \frac{1}{n}Sp\left( {\hat D_y^{\left( j \right)}} \right)\hat E + \Delta \hat D_y^{\left( j \right)}
\end{eqnarray}%
%---------------------------------------------------------------------------------
where $\hat E$ is unit matrix, the eigenvalues of the traceless matrix $\Delta \hat D_y^{\left( j \right)}$ will be small. Then
%---------------------------------------------------------------------------------
\begin{widetext}
\begin{equation}
\frac{1}{2}{\left( {\Delta \hat X_j^{\left( 0 \right)}} \right)^T}{{\hat D}^{\left( j \right)}}\Delta \hat X_j^{\left( 0 \right)}  
 =\frac{1}{n}Sp\left( {D_y^{\left( j \right)}} \right){\left| {{{\vec y}^{\left( 0 \right)}}} \right|^2}\left( {1 - \cos \left( {{\theta _j}} \right)} \right)
 + \frac{1}{2}\sum\limits_{k = 1}^n {\Delta d_{y,k}^{\left( j \right)}} {\left( {{V_{kn}}\left( {y_n^{\left( 0 \right)} - {{\hat P}_j}^{ - 1}\left( {y_n^{\left( 0 \right)}} \right)} \right)} \right)^2}
\label{eq10}
\end{equation}
\end{widetext}
%---------------------------------------------------------------------------------
where $\Delta d_{y,k}^{\left( j \right)}$ are the eigenvalues of matrix $\Delta \hat D_y^{\left( j \right)}$, ${V_{kn}}$ is the transformation matrix to the basis composed from the eigenvectors of matrix $\Delta \hat D_y^{\left( j \right)}$ (the summation over repeated indices is supposed). The second term in this sum is small in comparison with the first one due to the smallness of eigenvalues $\Delta d_{y,k}^{\left( j \right)}$ as well as due to their different signs (since the trace of matrix $\Delta \hat D_y^{\left( j \right)}$ is zero, the different terms over $k$ partially compensate each other). Therefore, we can adopt the following approximation:
%---------------------------------------------------------------------------------
\begin{eqnarray}%
&& \mbox{\fontsize{11}{14}\selectfont $ \frac{1}{2}{\left( {\Delta \hat X_j^{\left( 0 \right)}} \right)^T}{\hat D^{\left( j \right)}}\Delta \hat X_j^{\left( 0 \right)} \approx $}\nonumber\\
&& \mbox{\fontsize{11}{14}\selectfont $\approx \frac{1}{n}Sp\left( {\hat D_y^{\left( j \right)}} \right)  \times {\left| {{{\vec y}^{\left( 0 \right)}}} \right|^2}\left( {1 - \cos \left( {{\theta _j}} \right)} \right) $} \nonumber\\
\label{eq11}
\end{eqnarray}%
%---------------------------------------------------------------------------------

To approximately calculate the trace of matrix $\hat D_y^{\left( j \right)}$ we select the spherically symmetric part of matrix ${\hat D_y}$ representing it in the form 
%---------------------------------------------------------------------------------
\begin{eqnarray}%
&& {\hat D_y} = \frac{1}{n}Sp\left( {{{\hat D}_y}} \right)\hat E + \Delta {\hat D_y}
\label{eq11a}
\end{eqnarray}%
%---------------------------------------------------------------------------------
The results of numeral calculation of the eigenvalues of matrix ${\hat D_y}$ (which are denoted through $d_k^{\left( y \right)},k = 1,2, \cdots ,n$) are shown in Table.\ref{fig:part3_table01}. It is obvious that most eigenvalues are close between themselves with the exception of a few eigenvalues, which are substantially smaller. Therefore, these smallest eigenvalues have the highest absolute value of deviations from mean eigenvalue $\frac{1}{n}Sp\left( {{{\hat D}_y}} \right)$. Since all the eigenvalues of matrix ${\hat D_y}$ are positive, which means that their deviation from average value is less than this average in absolute value (see Table.\ref{fig:part3_table01}). Note that the matrix $\hat D_y^{\left( j \right)}$ can be represented in the following form:
%---------------------------------------------------------------------------------
\begin{widetext}
\begin{equation}
\hat D_y^{\left( j \right)} = \frac{1}{{2n}}Sp\left( {{{\hat D}_y}} \right)\left( {\hat E + \frac{{\Delta {{\hat D}_y}}}{{\frac{1}{n}Sp\left( {{{\hat D}_y}} \right)}}} \right)
 {\left( {\hat E + \frac{{\Delta {{\hat D}_y} + {{\hat P}_j}^T\Delta {{\hat D}_y}{{\hat P}_j}}}{{\frac{2}{n}Sp\left( {{{\hat D}_y}} \right)}}} \right)^{ - 1}}\left( {\hat E + \frac{{{{\hat P}_j}^T\Delta {{\hat D}_y}{{\hat P}_j}}}{{\frac{1}{n}Sp\left( {{{\hat D}_y}} \right)}}} \right) 
\label{eq12}
\end{equation}
\end{widetext}
%---------------------------------------------------------------------------------
By analogy we can conclude that the minimum eigenvalue of matrix 
%---------------------------------------------------------------------------------
\begin{eqnarray}%
&& \Delta {\hat D_y} + {\hat P_j}^T\Delta {\hat D_y}{\hat P_j}
\label{eq12a}
\end{eqnarray}%
%---------------------------------------------------------------------------------
(which is maximum in absolute value, see Table.\ref{fig:part3_table01}) is greater than the doubled minimum eigenvalue of matrix $\Delta {\hat D_y}$. This means that all the eigenvalues of matrix $\frac{{\Delta {{\hat D}_y} + {{\hat P}_j}^T\Delta {{\hat D}_y}{{\hat P}_j}}}{{\frac{2}{n}Sp\left( {{{\hat D}_y}} \right)}}$ are less than unity in absolute value. It applies equally to the eigenvalues of matrices $\frac{{\Delta {{\hat D}_y}}}{{\frac{1}{n}Sp\left( {{{\hat D}_y}} \right)}}$ and ${\hat P_j}^T\frac{{\Delta {{\hat D}_y}}}{{\frac{1}{n}Sp\left( {{{\hat D}_y}} \right)}}{\hat P_j}$. Therefore, we can represent the matrix $\hat D_y^{\left( j \right)}$ as the expansion in powers of $\frac{{\Delta {{\hat D}_y}}}{{\frac{1}{n}Sp\left( {{{\hat D}_y}} \right)}}$. Since matrix $\frac{{\Delta {{\hat D}_y}}}{{\frac{1}{n}Sp\left( {{{\hat D}_y}} \right)}}$ is traceless by definition,
 then a nonzero contribution to $Sp\left( {\hat D_y^{\left( j \right)}} \right)$ in addition to the term of ``zero'' order $\frac{1}{{2n}}Sp\left( {{{\hat D}_y}} \right)\hat E$ can give terms starting with the second-order. As it follows from Table.\ref{fig:part3_table01}, the maximum in absolute value eigenvalue of matrix $\frac{{\Delta {{\hat D}_y}}}{{\frac{1}{n}Sp\left( {{{\hat D}_y}} \right)}}$ increases with the energy growth. Therefore, we can expect that at ``low'' energies higher-order terms will make negligibly small contributions. In such an approximation we have:
%---------------------------------------------------------------------------------
\begin{eqnarray}%
&& Sp\left( {\hat D_y^{\left( j \right)}} \right) \approx \frac{1}{2}Sp\left( {{{\hat D}_y}} \right) 
\label{eq13}
\end{eqnarray}%
%---------------------------------------------------------------------------------
Let Eq.\ref{eq4} is taken in place of Eq.\ref{eq7} in approximation Eq.\ref{eq13}, then we have
%---------------------------------------------------------------------------------
\begin{eqnarray}%
& \mbox{\fontsize{11}{14}\selectfont $  \left\langle {{\sigma '_n}\left( {{z_l}} \right)} \right\rangle  = {\left( {A\left( {{{\hat X}^{\left( 0 \right)}}} \right)} \right)^2}v\left( {\sqrt s } \right) $} \nonumber\\ 
& \mbox{\fontsize{11}{14}\selectfont $ \times \exp \left( { - \frac{{{{\left| {{{\vec y}^{\left( 0 \right)}}} \right|}^2}Sp\left( {{{\hat D}_y}} \right)}}{{2n}}\left( {1 - {z_l}} \right)} \right) $}\nonumber\\ 
& \mbox{\fontsize{11}{14}\selectfont $ \times \frac{1}{{\Delta {N_l}}}\sum\limits_{{z_j} = {z_l}} {\frac{1}{{\sqrt {\det \left( {\frac{1}{2}\left( {\hat D + \hat P_j^T\hat D{{\hat P}_j}} \right)} \right)} }} } $ }
\label{eq14}
\end{eqnarray}%
%---------------------------------------------------------------------------------
Let us introduce the following notation
%---------------------------------------------------------------------------------
\begin{eqnarray}%
&& \mbox{\fontsize{12}{12}\selectfont $ \left\langle {w\left( {{z_l}} \right)} \right\rangle  = \frac{1}{{\Delta {N_l}}}\sum\limits_{{z_j} = {z_l}} {\frac{1}{{\sqrt {\det \left( {\frac{1}{2}\left( {\hat D + \hat P_j^T\hat D{{\hat P}_j}} \right)} \right)} }}}$ } 
\label{eq15}
\end{eqnarray}%
%---------------------------------------------------------------------------------
If we assume that multiplier $\left\langle {w\left( {{z_l}} \right)} \right\rangle $ is weakly dependent on ${z_l}$, we obtain
%---------------------------------------------------------------------------------
\begin{eqnarray}%
&& \left\langle {{\sigma '_n}\left( {{z_l}} \right)} \right\rangle  = \left\langle {{\sigma '_n}\left( {{z_{{l_0}}}} \right)} \right\rangle   \exp \left( {\frac{{{{\left| {{{\vec y}^{\left( 0 \right)}}} \right|}^2}Sp\left( {{{\hat D}_y}} \right)}}{{2n}}\left( {{z_l} - {z_{{l_0}}}} \right)} \right) \nonumber\\ 
\label{eq16}
\end{eqnarray}%
%---------------------------------------------------------------------------------
where ${z_{l_{0}}}$ is the minimum value of ${z_l}$ for which can be numerically calculated all interference contributions. Therefore, the magnitude $\left\langle {{\sigma '_n}\left( {{z_{{l_0}}}} \right)} \right\rangle $ can be directly calculated numerically. The results of numerical calculation of $\ln \left( {\left\langle {{\sigma '_n}\left( {{z_l}} \right)} \right\rangle } \right)$ over all interference contributions in comparison with the results obtained by Eq.\ref{eq15} are demonstrated on Fig.\ref{fig:part3_fig04}, it follows that such an approximation is acceptable at ``low'' energies.

Results shown in Fig.\ref{fig:part3_fig04} confirm also our assumption that $\left\langle {w\left( {{z_l}} \right)} \right\rangle $ weakly depends on ${z_l}$. To analyze this dependence we turn to Fig.\ref{fig:part3_fig05}. It is obvious, that the magnitude $\left\langle {w\left( {{z_l}} \right)} \right\rangle $ takes small values at ``low'' energies. This means that 
%---------------------------------------------------------------------------------
\begin{eqnarray}%
&& \det \left( {\frac{1}{2}\left( {\hat D + \hat P_j^T\hat D{{\hat P}_j}} \right)} \right)
\label{eq_determinant}
\end{eqnarray}%
%---------------------------------------------------------------------------------
% takes large values at the same energies. Indeed, as it follows from the expression for the matrix $\hat D$, Eq.\ref{eq_determinant} tends to infinity on the threshold of $n$ particle production, and this means that at threshold the phase space of physical area of the inelastic process with $n$ particles production takes place is equal to zero.
takes large values at the same energies. Indeed, as it follows from the expression for the matrix $\hat D$, Eq.\ref{eq_determinant}tends to infinity on the threshold of $n$ particles production, and this means that at the threshold the volume of phase space with $n$ particles production in the inelastic process is equal to zero. 
%%%%%%%%%%%%---------------------------------%%%%%%%%%
%%%%%%%%%%%%---------------------------------%%%%%%%%%
\begin{figure}
\begin{center}
  \includegraphics[scale=0.24]{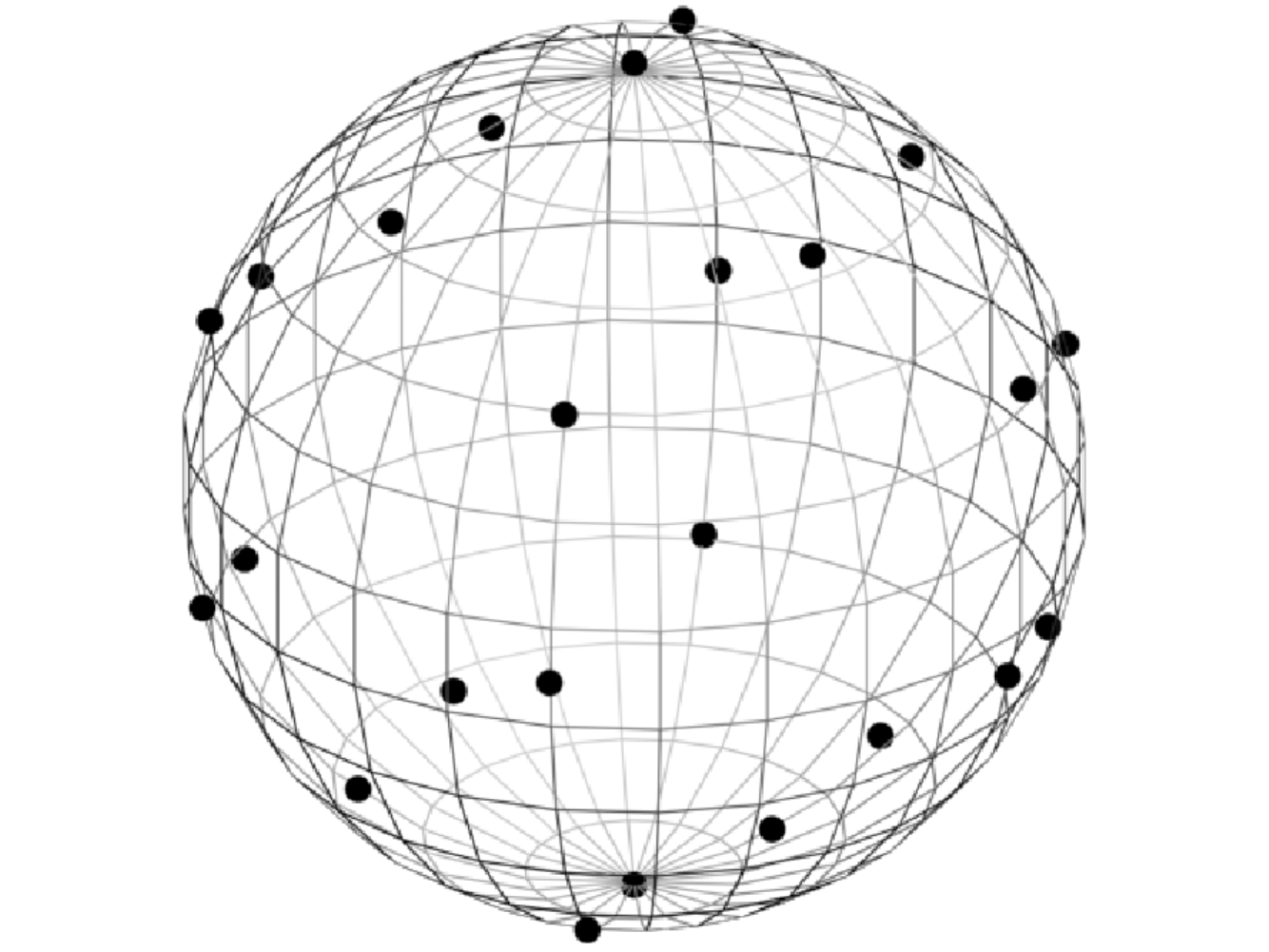} 
\end{center}
 \caption{A sphere ${S_2}$ and figure ${F_{4!}}$, which is demonstrated by points. Basis in the four-dimensional space is chosen so that the one of vectors coincides with the vector ${\vec e_4} = \left( {\frac{1}{2},\frac{1}{2},\frac{1}{2},\frac{1}{2}} \right)$, and the three basis vectors of three-dimensional subspace, into which depicted sphere is embedded, are perpendicular to ${\vec e_4}$. }
 \label{fig:part3_fig06}
\end{figure}
%%%%%%%%%%%%---------------------------------%%%%%%%%%
%%%%%%%%%%%%---------------------------------%%%%%%%%%
%%%%%%%%%%%%---------------------------------%%%%%%%%%
%%%%%%%%%%%%---------------------------------%%%%%%%%%
\begin{figure}
\begin{center}
  \centering
%  \subfigure[]{
%  \includegraphics[scale=0.263]{fig_part3_06a} 
%  \label{fig:fig_part3_06a} 
%  }
  \subfigure[]{
  \includegraphics[scale=0.153]{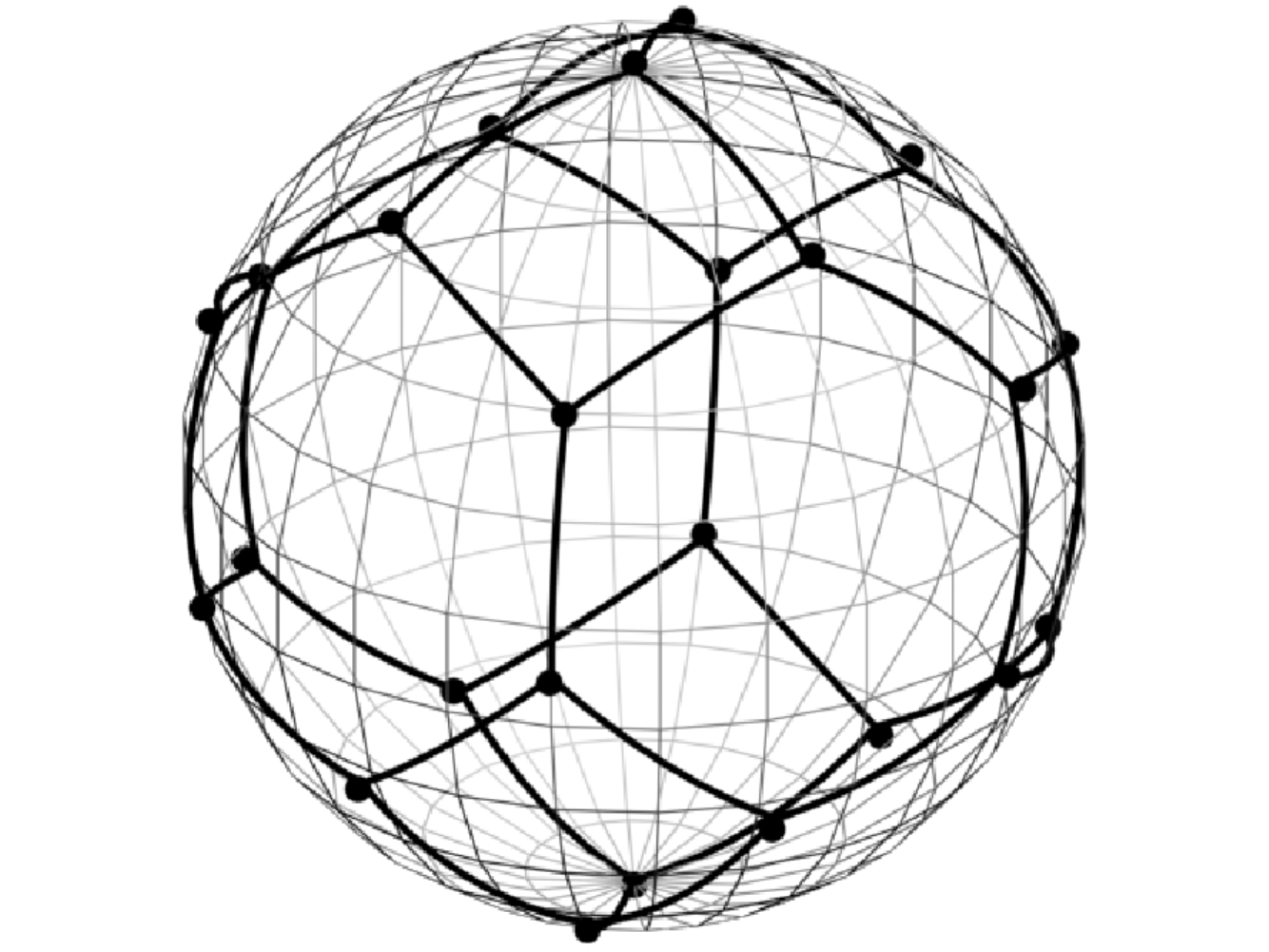} 
  \label{fig:part3_fig07a} 
  }
  \subfigure[]{
  \includegraphics[scale=0.153]{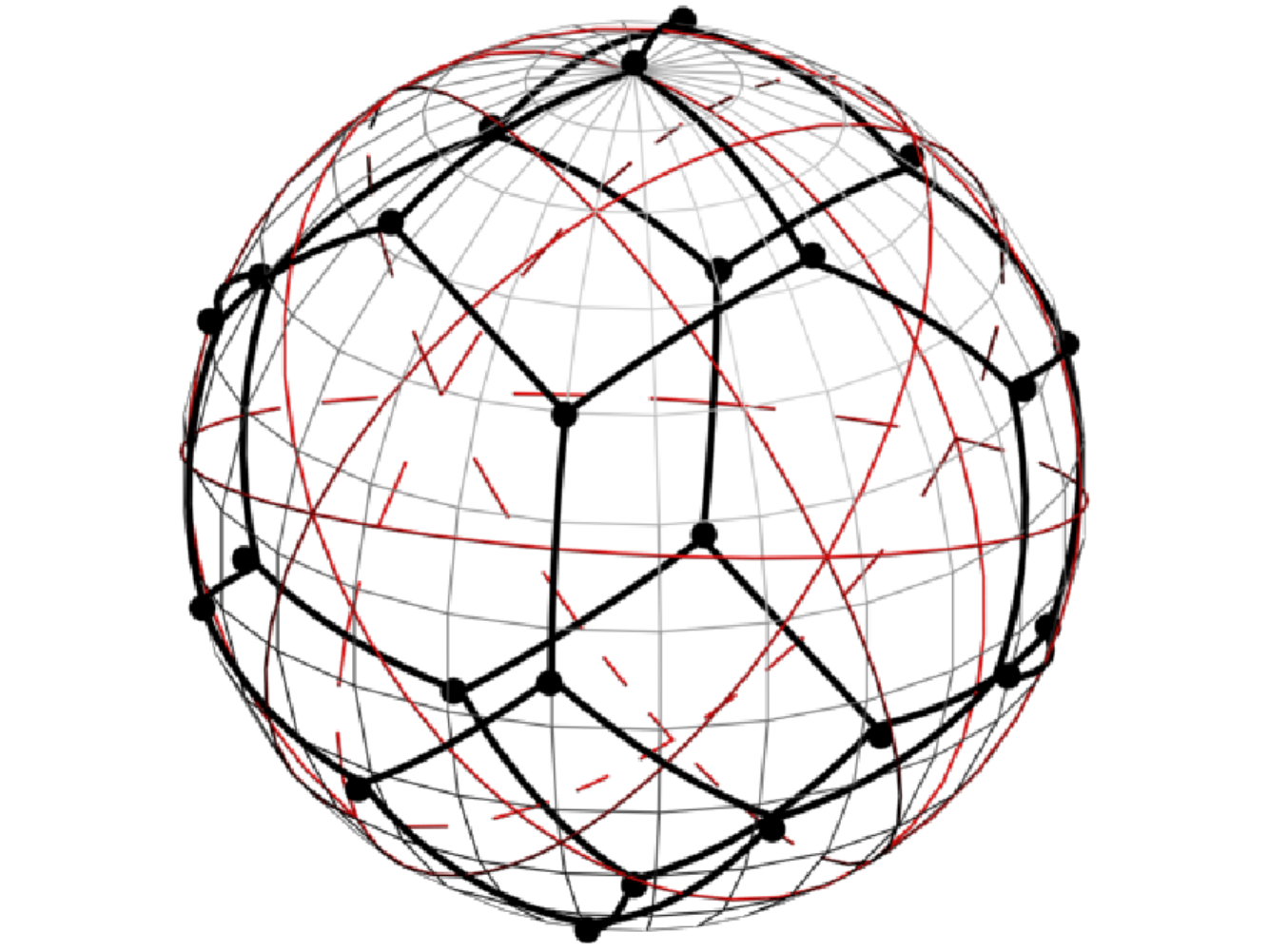} 
  \label{fig:part3_fig07b} 
  }
  \subfigure[]{
  \includegraphics[scale=0.153]{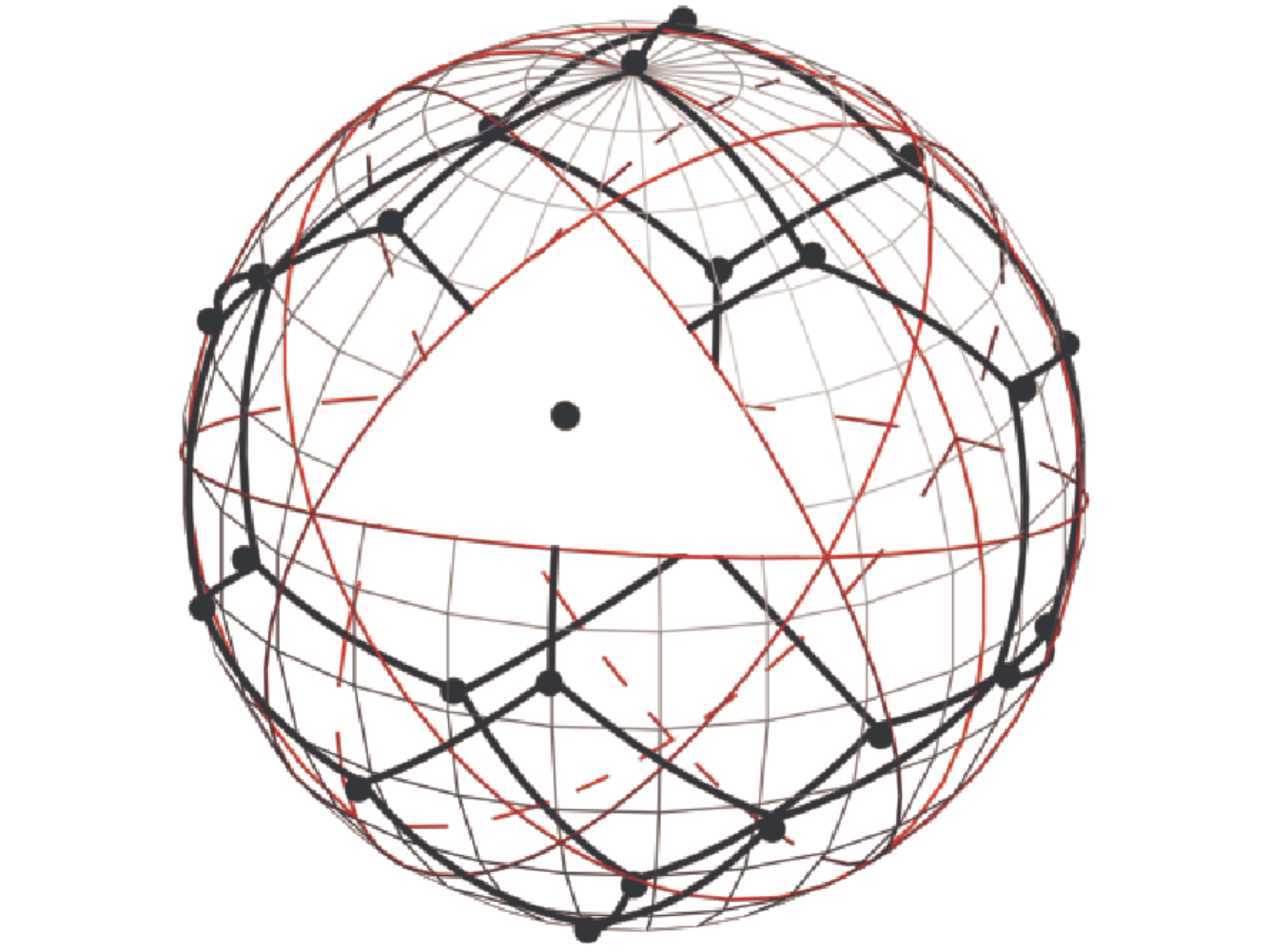} 
  \label{fig:part3_fig07c} 
  } 
\end{center}
 \caption{%The partition of sphere ${S_2}$ by shortest arcs joining the points of figure ${F_{4!}}$ into the two ``hexagonal'' and one ``tetragonal'' regions \ref{fig:part3_fig07a}; (b) areas, which is located on the borders of 4 or 6 points belonging to figure ${F_{4!}}$ can be divided between those points into figures of equal area; (c) whole sphere $S_2$ is divided into figures of equal area, each of which contains the one point of figure ${F_{4!}}$ one of these shapes are painted in white.}
Sphere ${S_2}$ partitioning by shortest arcs joining the points of figure ${F_{4!}}$ into the two ``hexagonal'' and one ``tetragonal'' regions \ref{fig:part3_fig07a}; (b) areas, which is located on the borders of 4 or 6 points belonging to figure ${F_{4!}}$ can be divided between those points into figures of equal area; (c) whole sphere $S_2$ is divided into figures of equal area, each of which contains the one point of figure ${F_{4!}}$ $\textendash$ one of those shapes are painted in white.}
 \label{fig:part3_fig07}
\end{figure}
%%%%%%%%%%%%---------------------------------%%%%%%%%%
%%%%%%%%%%%%---------------------------------%%%%%%%%%

Because of symmetry with respect to direction inversion in a plane of transversal momenta the mixed second derivatives with respect to rapidities and transversal momentum components are zeros. As a consequence, the determinant Eq.\ref{eq_determinant} is equal to the product of the three determinants, first of which is composed from second derivatives with respect to rapidities, the second is composed from the second derivatives with respect to the transversal momentum $x$-components and the third one is composed from derivatives with respect to the transversal momentum $y$-components. All the three factors tend to infinity at the threshold energy. As it follows from a numerical calculation, a matrix determinant composed from the second derivatives with respect to rapidities reduced quite rapidly with energy growth. Matrix determinants composed from the second derivatives with respect to transversal momentum components also reduced, but in a wide energy range, they remain quite large. Therefore, the value of Eq.\ref{eq_determinant} is great at all $j$. Since the function ${1 \mathord{\left/
 {\vphantom {1 {\sqrt x }}} \right.
 \kern-\nulldelimiterspace} {\sqrt x }}$ varies slightly at the great values of argument, the function $\left\langle {w\left( {{z_l}} \right)} \right\rangle $ weakly depends on ${z_l}$. 
 
To estimate roughly the function $\left\langle {w\left( {{z_l}} \right)} \right\rangle $ we can replace it by the Taylor expansion taking into account just linear contributions. The expansion coefficients are found by the calculating of $\left\langle {w\left( {{z_l}} \right)} \right\rangle $ for ${z_l}$ close to $1$ and $(-1)$. In these cases the values of
%---------------------------------------------------------------------------------
\begin{eqnarray}%
&& {1/\sqrt {\det \left( {1/2\left( {\hat D + \hat P_j^T\hat D{{\hat P}_j}} \right)} \right)} }
\label{eq_detfraction}
\end{eqnarray}%
%---------------------------------------------------------------------------------
were obtained directly for all proper interference contributions, and after that we obtain the values of $\left\langle {w\left( {{z_l}} \right)} \right\rangle $ by averaging using Eq.\ref{eq15}.

The values in Fig.\ref{fig:part3_fig05} have been obtained by the direct calculation of
%---------------------------------------------------------------------------------
\begin{eqnarray}%
 && \mbox{\fontsize{11}{14}\selectfont $ \frac{1}{{\Delta {N_l}}}\sum\limits_{{z_j} = {z_l}} {1/\sqrt {\det \left( {1/2\left( {\hat D + \hat P_j^T\hat D{{\hat P}_j}} \right)} \right)} } $}
\end{eqnarray}%
%---------------------------------------------------------------------------------
with consideration of all interference contributions at different $\sqrt s $.

So, we have the following expression instead of Eq.\ref{eq16}
%---------------------------------------------------------------------------------
\begin{widetext}
\begin{equation}
 \left\langle {{\sigma '_n}\left( {{z_l}} \right)} \right\rangle  = \left\langle {{\sigma '_n}\left( {{z_{{l_0}}}} \right)} \right\rangle \left( {{w_0} + {w_1}\left( {1 - {z_l}} \right)} \right)
 \times \exp \left( {\frac{{{{\left| {{{\vec y}^{\left( 0 \right)}}} \right|}^2}Sp\left( {{{\hat D}_y}} \right)}}{{2n}}\left( {{z_l} - {z_{{l_0}}}} \right)} \right) 
\label{eq17}
\end{equation}
\end{widetext}
%---------------------------------------------------------------------------------
where the coefficients ${w_0}$ and ${w_1}$ found by above mentioned method.

%#################################################################
\section{Approximate calculation of the $\Delta {N_l}$ values}
\label{SECTION_4}
%#################################################################
Let us turn to the new variables
%---------------------------------------------------------------------------------
\begin{eqnarray}%
&& Y_k^{\left( 0 \right)} = \frac{{y_k^{\left( 0 \right)}}}{{\Delta y\left( {n, \sqrt s} \right)\sqrt {\frac{{\left( {n + 1} \right)n\left( {n - 1} \right)}}{{12}}} }} 
\label{eq18}
\end{eqnarray}%
%---------------------------------------------------------------------------------
where $y_k^{\left( 0 \right)}$ are determined by Eq.\ref{eq2}, $Y_k^{\left( 0 \right)},k = 1,2, \cdots ,n$ are considered as the components of vector ${\vec Y^{\left( 0 \right)}}$, which, as it follows from Eq.\ref{eq18} is of unit length.

Thus, the angle ${\theta _j}$ between the vector ${\vec y^{\left( 0 \right)}} = \left( {y_1^{\left( 0 \right)},y_2^{\left( 0 \right)}, \cdots ,y_n^{\left( 0 \right)}} \right)$ and vector $\hat P_j^{ - 1}\left( {{{\vec y}^{\left( 0 \right)}}} \right)$ obtained by the permutation of corresponding components is the same as the angle between the vector ${\vec Y^{\left( 0 \right)}} = \left( {Y_1^{\left( 0 \right)},Y_2^{\left( 0 \right)}, \cdots ,Y_n^{\left( 0 \right)}} \right)$ and vector $\hat P_j^{ - 1}\left( {{{\vec Y}^{\left( 0 \right)}}} \right)$. 
Moreover, as it follows from Eq.\ref{eq2}
%---------------------------------------------------------------------------------
\begin{eqnarray}%
&& y_1^{\left( 0 \right)} =  - y_n^{\left( 0 \right)},y_2^{\left( 0 \right)} =  - y_{n - 1}^{\left( 0 \right)}, \cdots ,
 y_k^{\left( 0 \right)} =  - y_{n - k + 1}^{\left( 0 \right)};\nonumber\\ 
&& k = 1,2, \cdots ,n 
\label{eq19}
\end{eqnarray}%
%---------------------------------------------------------------------------------
It follows that all vectors $\hat P_j^{-1} \left( \vec Y^{(0)} \right)$ are orthogonal to vector
%---------------------------------------------------------------------------------
\begin{eqnarray}%
&& \vec e_n = \left( \underbrace{1/\sqrt n,1/\sqrt n,\ldots,1/\sqrt n }_{n \quad \scriptsize{components}} \right)
 \label{eq19a}
 \end{eqnarray}%
 %---------------------------------------------------------------------------------
%%%%%%%%%%%%---------------------------------%%%%%%%%%
\begin{figure}
\begin{center}
  \includegraphics[scale=0.53]{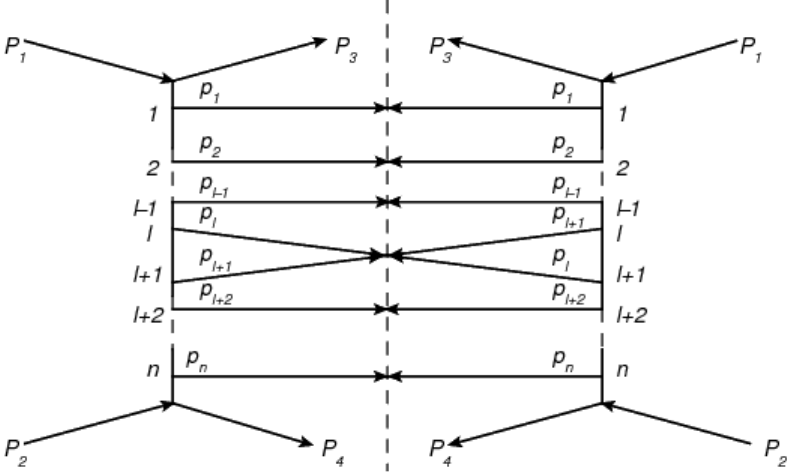} 
\end{center}
 \caption{Diagrams, which correspond $(n-1)$ vectors $\hat P_l^{ - 1}\left( {{{\vec Y}^{\left( 0 \right)}}} \right)$ closest to vector ${\vec Y^{\left( 0 \right)}}$. }
 \label{fig:part3_fig08}
\end{figure}
%%%%%%%%%%%%---------------------------------%%%%%%%%%
%%%%%%%%%%%%---------------------------------%%%%%%%%%
\begin{figure*}
\begin{center}
  \centering
  \subfigure[]{
  \includegraphics[scale=0.3]{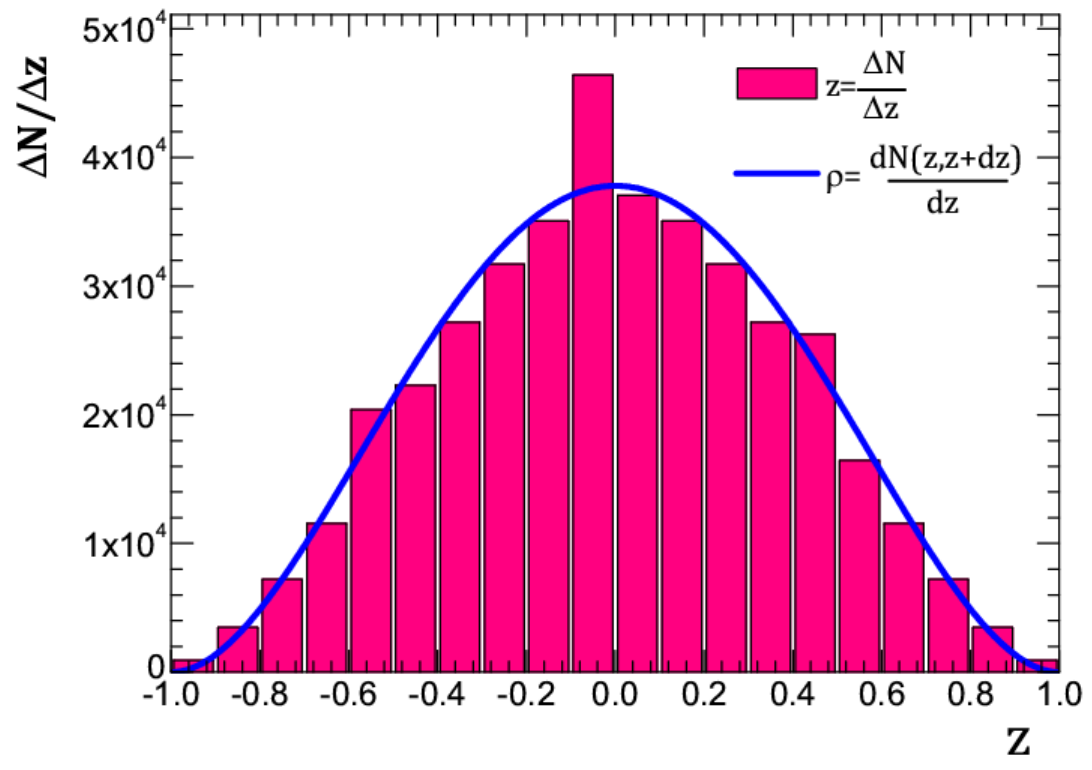} 
  \label{fig:part3_fig09a} 
  }
  \subfigure[]{
  \includegraphics[scale=0.3]{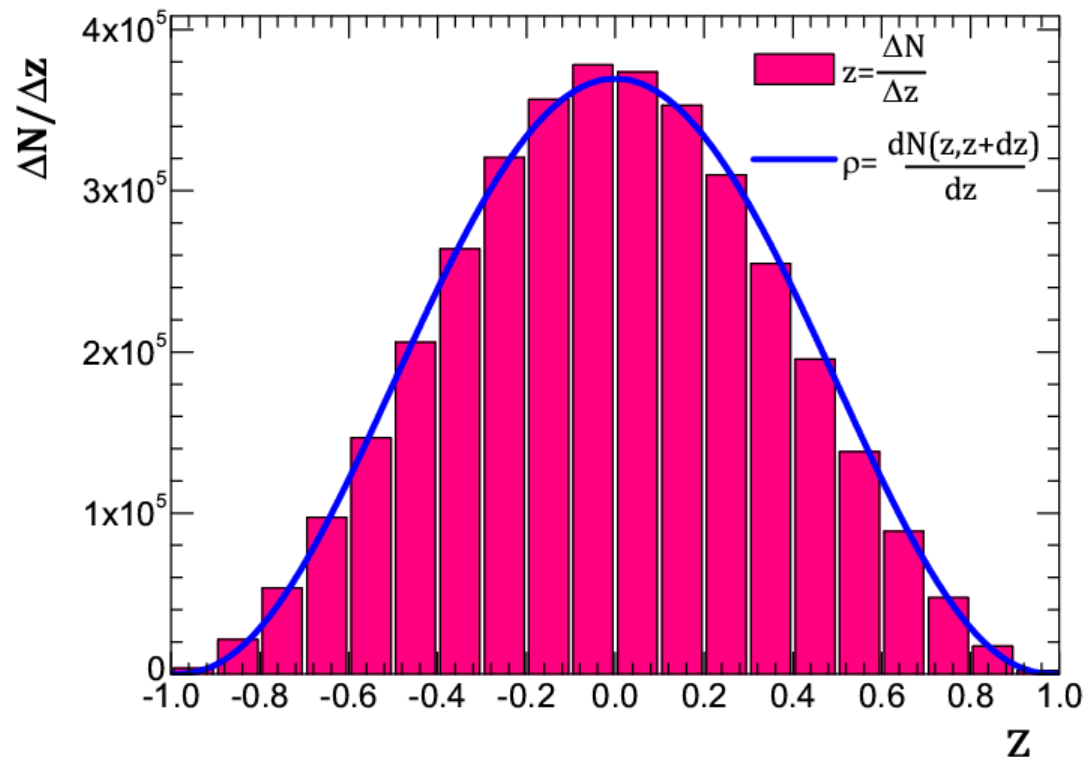} 
  \label{fig:part3_fig09b} 
  }
  \subfigure[]{ 
 \includegraphics[scale=0.3]{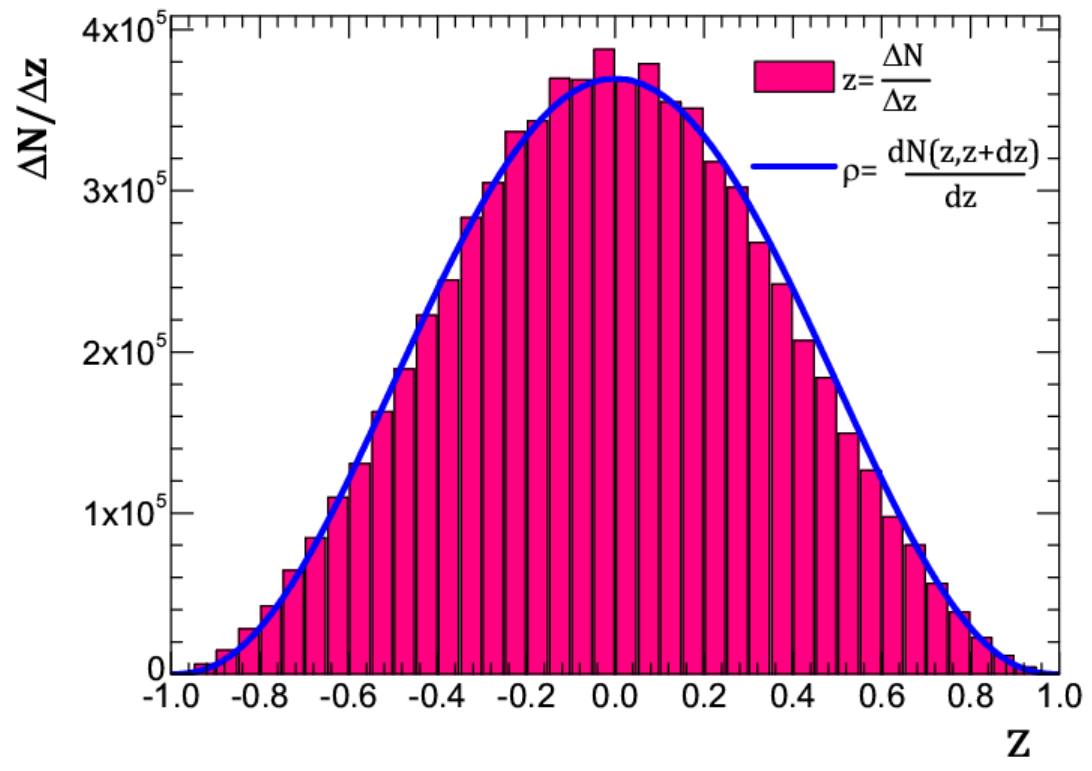} 
  \label{fig:part3_fig09c} 
  }
\end{center}
 \caption{Distribution of the interference contributions by the variable $z = \cos \left( \Theta  \right)$ (histogram) in comparison with the function $\rho \left( z \right) = \frac{{dN\left( {z,z + dz} \right)}}{{dz}}$ (solid line) at $n=8$,  $\Delta z = 0.1$ (a); $n=9$,  $\Delta z = 0.1$ (b); $n=9$,  ${\Delta z} = 0.05$ (c). Where  $\Delta N$ is the number of interference contributions corresponding to value of $z$ in the proper interval of  $\Delta z$ width. }
  \label{fig:part3_fig09} 
\end{figure*}
%%%%%%%%%%%%---------------------------------%%%%%%%%%

Therefore, considering vectors $\hat P_j^{-1} \left( \vec Y^{(0)} \right)$ as the elements of $n$-dimensional euclidean space, which we denote through $E_n$, then the ends of all vectors $\hat P_j^{-1} \left( \vec Y^{(0)} \right)$ are lie on the unit sphere embedded into the $(n-1)$-dimensional subspace of $E_n$. We denote this sphere through $S_{n-2}$ and shape formed by the set of points in which the ends of vectors $\hat P_j^{-1} \left( \vec Y^{(0)} \right)$ ($j = 1,2,\ldots,n!$) come, denote through $F_{n!}$. In particular, when $n=4$ the sphere $S_2$ and figure $F_{4!}$ graphically look like in Fig.\ref{fig:part3_fig06}.
%%%%%%%%%%%%---------------------------------%%%%%%%%%
\begin{figure*}
\begin{center}
  \centering
  \subfigure[]{
  \includegraphics[scale=0.38]{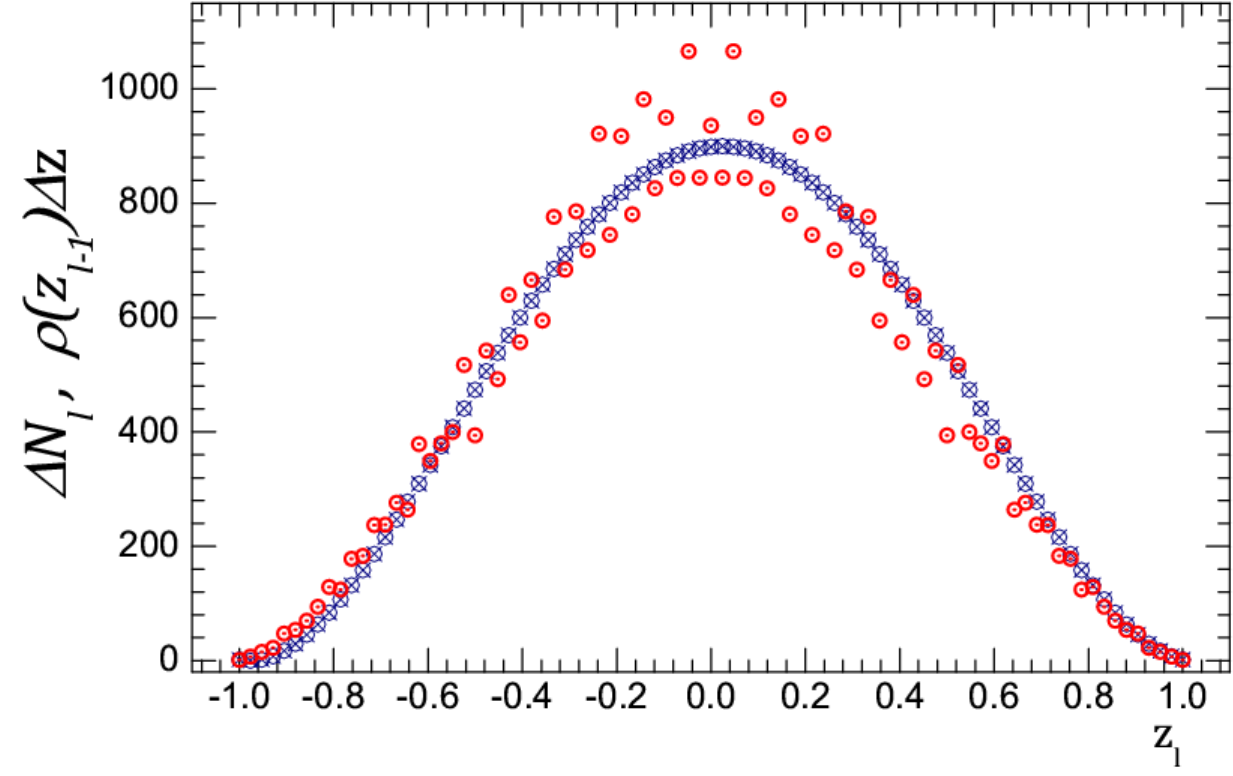} 
  \label{fig:part3_fig10a} 
  }
  \subfigure[]{
  \includegraphics[scale=0.38]{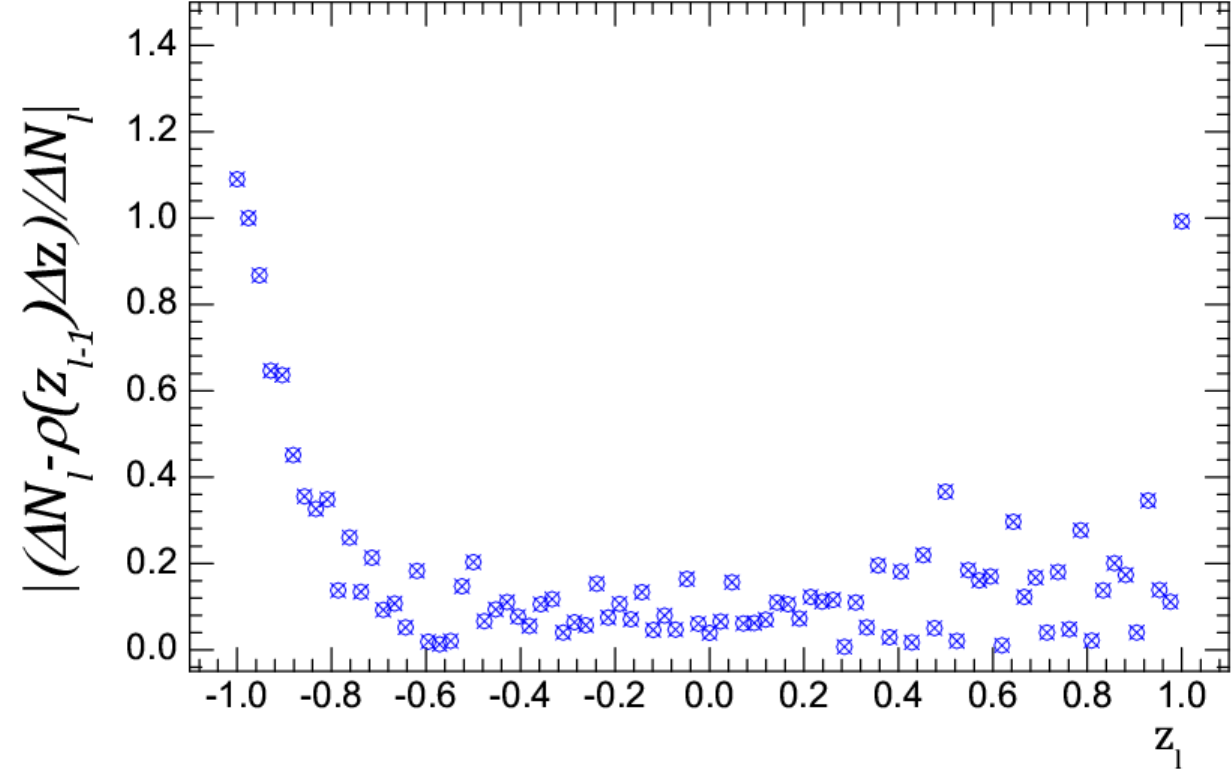} 
  \label{fig:part3_fig10b} 
  }
  \subfigure[]{
  \includegraphics[scale=0.38]{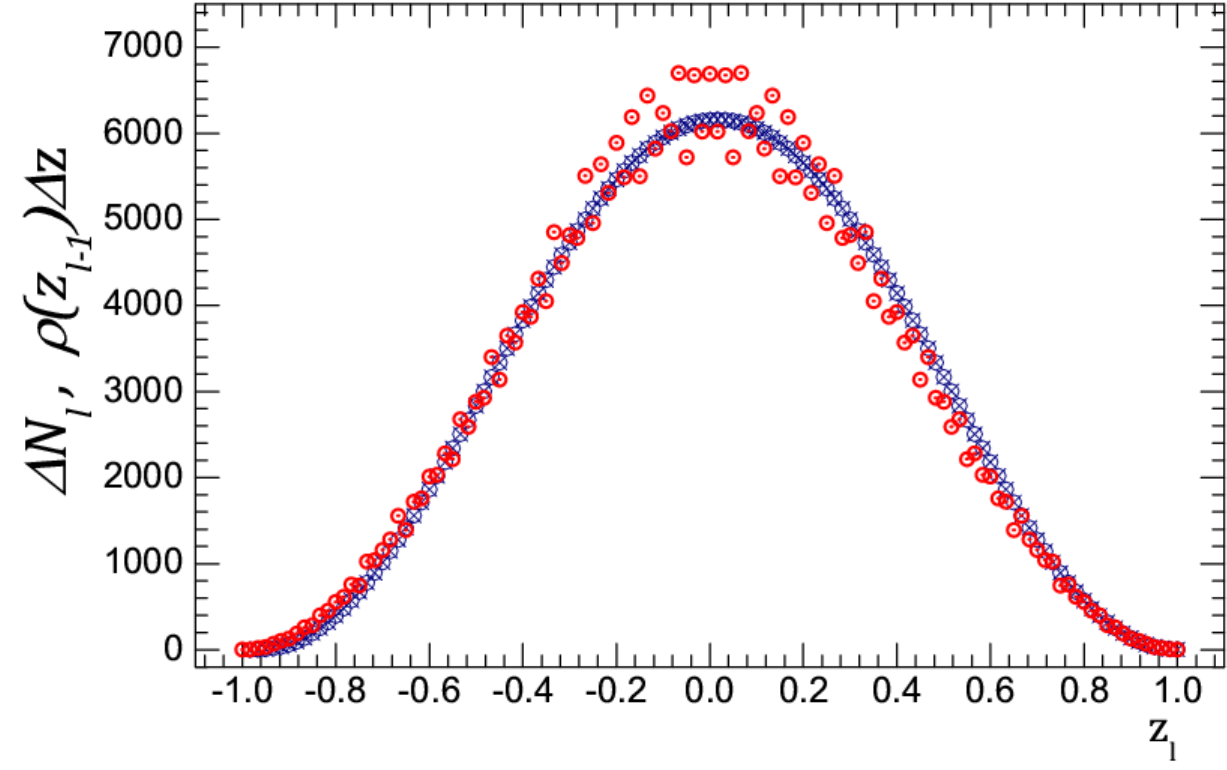} 
  \label{fig:part3_fig10c} 
  }
  \subfigure[]{
  \includegraphics[scale=0.38]{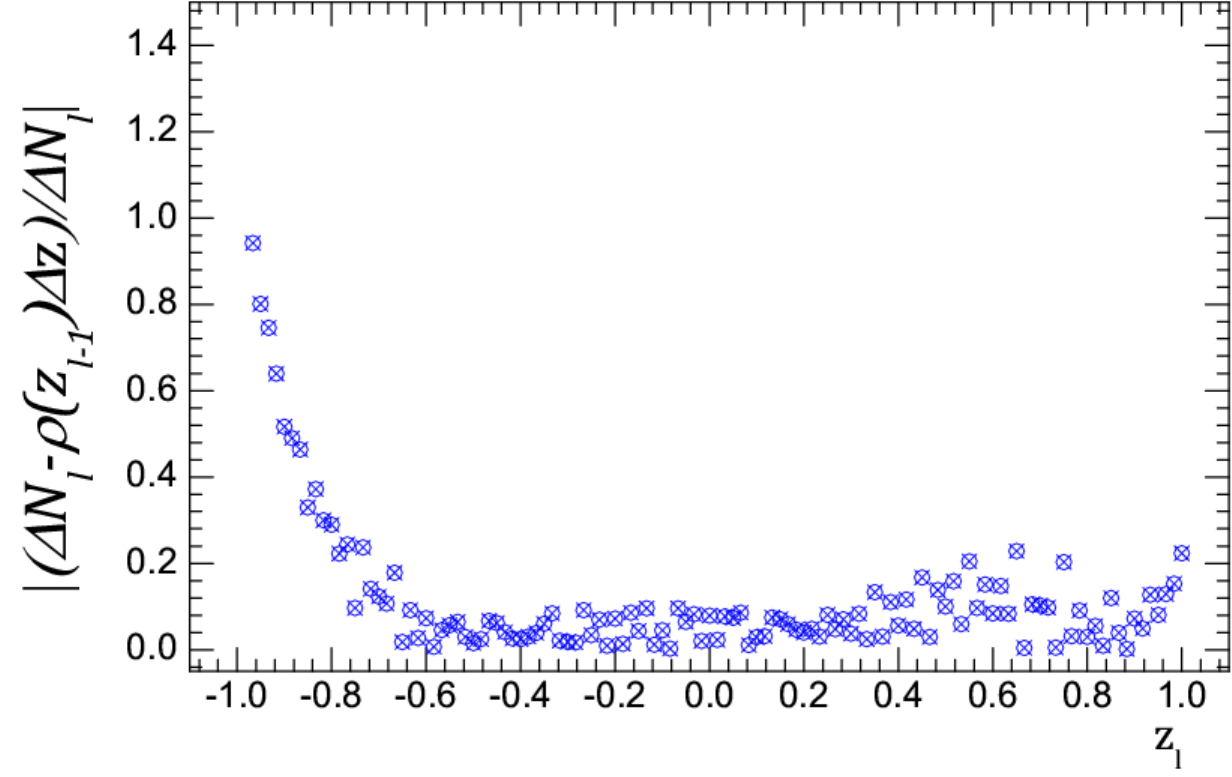} 
  \label{fig:part3_fig10d} 
  }
\end{center}
 \caption{
Comparison of the values of right-hand side and left-hand side of approximate equality Eq.\ref{eq23} at $n=8$ (a, b) and $n=9$ (c, d). Circles are the values of $\Delta {N_l}$ obtained with consideration of all interference contributions; crosses $\textendash$ are the values of function $\rho \left( {{z_{l - 1}}} \right)\Delta z$ from Eq.\ref{eq22}.}
  \label{fig:part3_fig10} 
\end{figure*}
%%%%%%%%%%%%---------------------------------%%%%%%%%%

We examine some geometrical properties of figure ${F_{n!}}$ at arbitrary $n$. If we apply the permutation transformation component to all vectors in the $n$-dimensional space, where the vectors ${\vec Y^{\left( 0 \right)}}$ are primordially defined, the examined $(n-1)$-dimensional subspace as well as a sphere ${S_{n - 2}}$ and figure ${F_{n!}}$ go into themselves. As it follows from the group properties of permutation group, the each point of figure ${F_{n!}}$ can be obtained from any other point by some transformation $\hat P_j^{ - 1}$. This means, that the configuration of the points of figure ${F_{n!}}$ relative to each of these points must be identical, that can be clearly seen in Fig.\ref{fig:part3_fig07a}.

As it follows from Eq.\ref{eq19}, besides the end of each vector $\hat P_j^{ - 1}\left( {{{\vec Y}^{\left( 0 \right)}}} \right)$ a figure ${F_{n!}}$ contains also the end of vector $\left( { - \hat P_j^{ - 1}\left( {{{\vec Y}^{\left( 0 \right)}}} \right)} \right)$, i.e., a figure ${F_{n!}}$ has a center of symmetry, which coincides with the center of sphere ${S_{n - 2}}$. In this case, if we using point of ${F_{n!}}$ form path from the point $\hat P_j^{ - 1}\left( {{{\vec Y}^{\left( 0 \right)}}} \right)$ to the point $\left( { - \hat P_j^{ - 1}\left( {{{\vec Y}^{\left( 0 \right)}}} \right)} \right)$, then it will be simultaneously formed a centro-symmetrical path, that leads from $\left( { - \hat P_j^{ - 1}\left( {{{\vec Y}^{\left( 0 \right)}}} \right)} \right)$ to $\hat P_j^{ - 1}\left( {{{\vec Y}^{\left( 0 \right)}}} \right)$ of figure ${F_{n!}}$. 

Joining these paths we will obtain the closed path, which ``girdles'' the sphere ${S_{n - 2}}$. If we assume that there is such a ``girdling'' path, inside of which are concentrated all points of figure ${F_{n!}}$, we would find that the figure ${F_{n!}}$ has a ``boundary'' and ``internal'' points, that would contradict the fact that spacing of all points relative to each point of the ${F_{n!}}$ should be the same. In other words, the points of figure ${F_{n!}}$ must ``crawl away'' all over the sphere ${S_{n - 2}}$ and can not be concentrated on some area of the sphere.

If we consider a vector ${\vec Y^{\left( 0 \right)}}$, then vectors corresponding to permutations $\hat P_l^{ - 1},l = 1,2, \cdots ,n - 1$ defined by the following relation
%---------------------------------------------------------------------------------
\begin{eqnarray}%
 && \mbox{\fontsize{11}{14}\selectfont $  {\left( {\hat P_l^{ - 1}\left( {{{\vec Y}^{\left( 0 \right)}}} \right)} \right)_k} = \left\{ \begin{array}{l}
 Y_k^{\left( 0 \right)},{\rm{if }}\quad k < l, \\ 
 Y_{l + 1}^{\left( 0 \right)},{\rm{if }}\quad k = l, \\ 
 Y_l^{\left( 0 \right)},{\rm{if }}\quad k = l + 1, \\ 
 Y_k^{\left( 0 \right)},{\rm{if }}\quad k > l + 1. \\ 
 \end{array} \right. $}
 \label{eq20}
\end{eqnarray}%
%---------------------------------------------------------------------------------
will be closest to it. The type of ``cut'' diagrams corresponding to such permutations is shown on Fig.\ref{fig:part3_fig08}.

At the same time, all the components of vector $\hat P_l^{ - 1}\left( {{{\vec Y}^{\left( 0 \right)}}} \right) - {\vec Y^{\left( 0 \right)}}$ , except the $l$-th and $l + 1$, are zero, whereas these two components take on the least values in modulus $\sqrt {\frac{{12}}{{\left( {n + 1} \right)n\left( {n - 1} \right)}}} $ and $\left( { - \sqrt {\frac{{12}}{{\left( {n + 1} \right)n\left( {n - 1} \right)}}} } \right)$, respectively.

Thus, we can conclude that the each point of figure ${F_{n!}}$ has $(n-1)$ nearest neighboring points, which lying at distance of from it:
%---------------------------------------------------------------------------------
\begin{eqnarray}%
&& {r_n} = \sqrt {\frac{{24}}{{\left( {n + 1} \right)n\left( {n - 1} \right)}}} 
\label{eq21} 
\end{eqnarray}%
%---------------------------------------------------------------------------------
Connecting the each point of figure ${F_{n!}}$ with its (n-1) nearest neighbors points by shortest arc, thereby we divide the sphere ${S_{n - 2}}$ into closed regions as is shown in Fig.\ref{fig:part3_fig07a}. Indeed, let us choose the some point ${A_0}$ of figure ${F_{n!}}$, and will move from it to the nearest point $A_1$ along a shortest arc, then we move from the point $A_1$ to the nearest point $A_2$ etc. At the same time, motion in a backward direction is prohibited. Thus, there are $(n-1)$ paths going out from each point, and $(n-2)$ paths are allowed at each step. But since figure ${F_{n!}}$ has the final number of points at some step we will surely come back to the point ${A_0}$.

Moreover, since shortest arcs joining two nearest points are subtended by equal chords ${r_n}$ in length (see Eq.\ref{eq21}), this arcs are of the same length. Let us consider any two neighboring points ${A_i}$ and ${A_{i + 1}}$ of figure ${F_{n!}}$. Under any transformation $\hat P_j^{ - 1}$ the shortest arc, which joins the points ${A_i}$ and ${A_{i + 1}}$, and an arc joining the points $\hat P_j^{ - 1}\left( {{A_i}} \right)$ and $\hat P_j^{ - 1}\left( {{A_{i + 1}}} \right)$ are of the same length. This means that the boundaries of closed regions formed by shortest arcs, which join neighboring points, replaced into one another under any transformation $\hat P_j^{ - 1}$. It follows that, if we examine closed regions, which include any point of figure ${F_{n!}}$, then the adjacent areas to all points of this figure will have the same ``area''.
%%%%%%%%%%%%---------------------------------%%%%%%%%%
%%%%%%%%%%%%---------------------------------%%%%%%%%%
\begin{figure*}
\begin{center}
  \centering
  \subfigure[]{
  \includegraphics[scale=0.38]{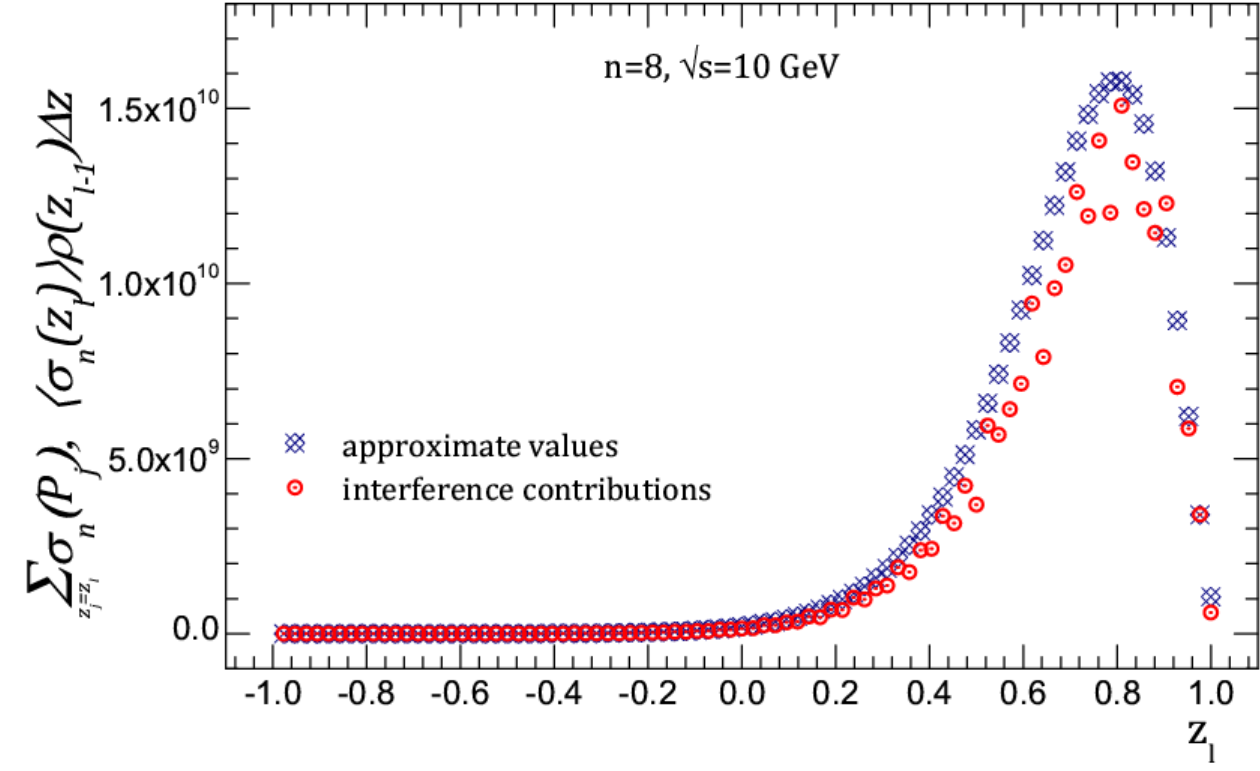} 
  \label{fig:part3_fig11a} 
  }
  \subfigure[]{
  \includegraphics[scale=0.38]{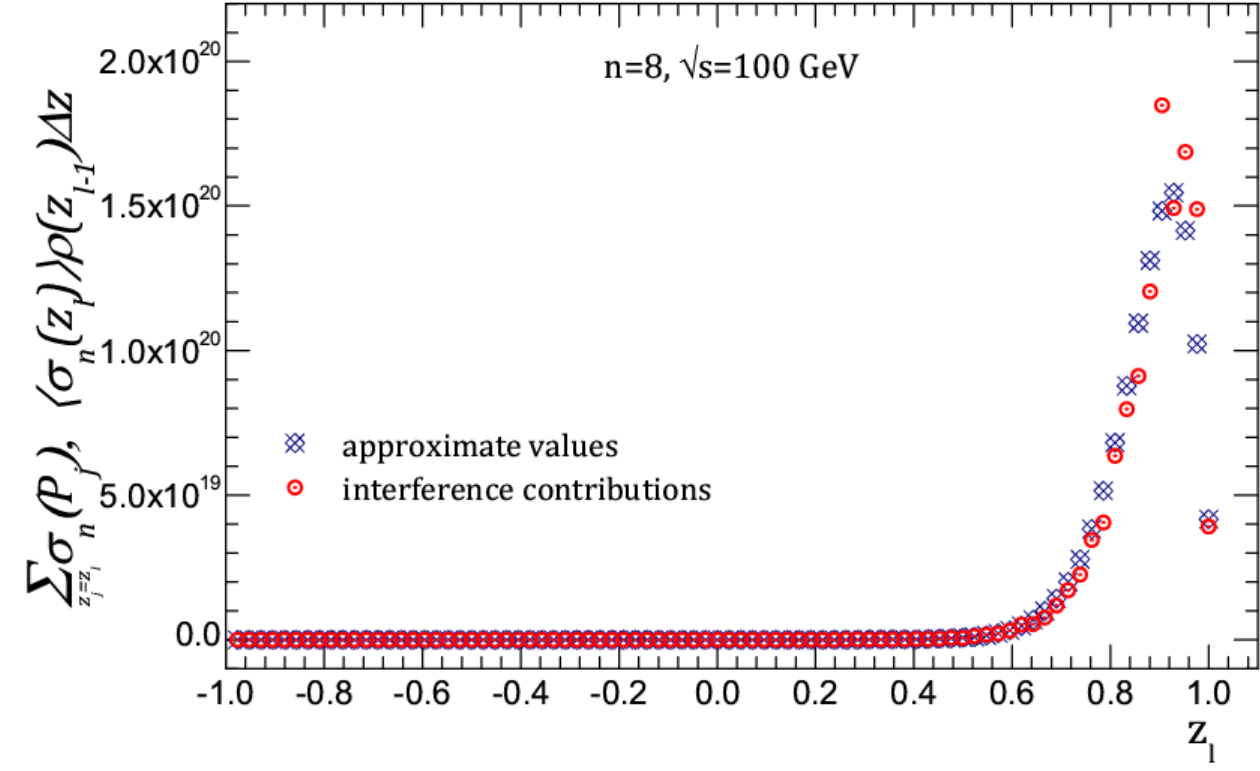} 
  \label{fig:part3_fig11b} 
  }
  \subfigure[]{ 
  \includegraphics[scale=0.38]{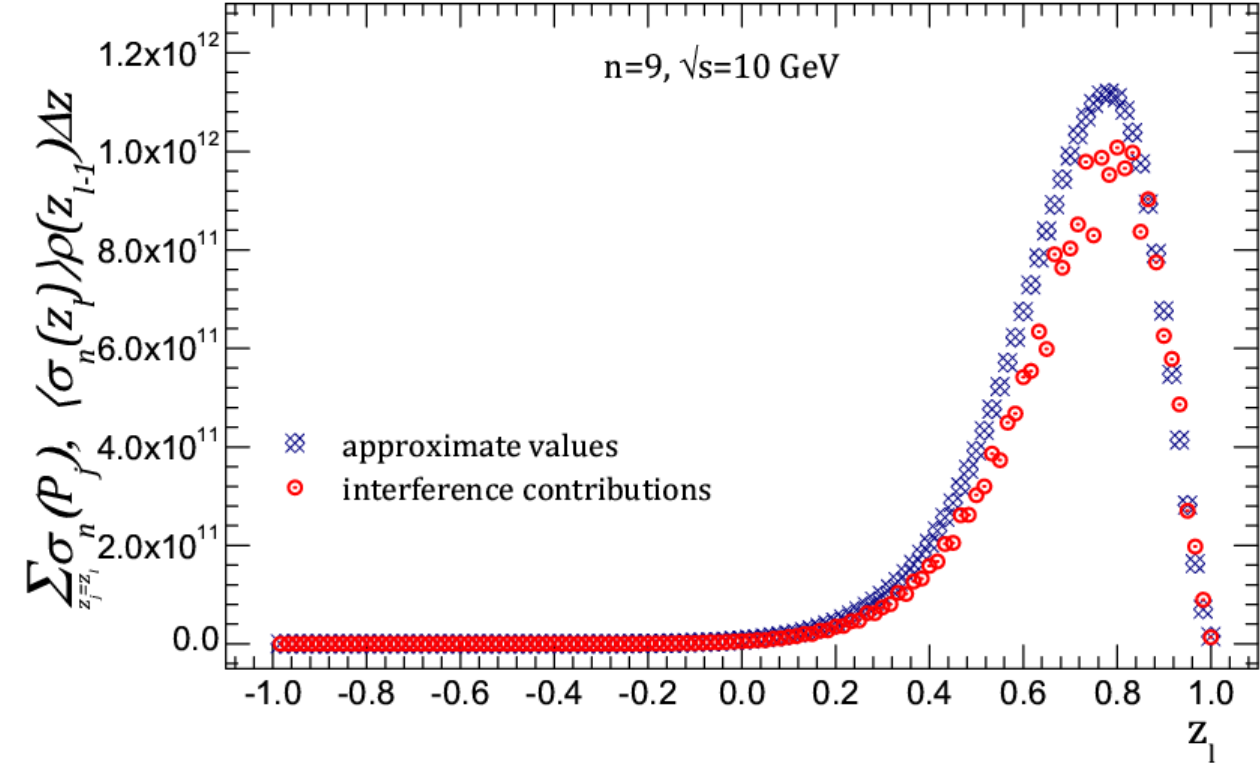} 
  \label{fig:part3_fig11c} 
  }
  \subfigure[]{ 
  \includegraphics[scale=0.38]{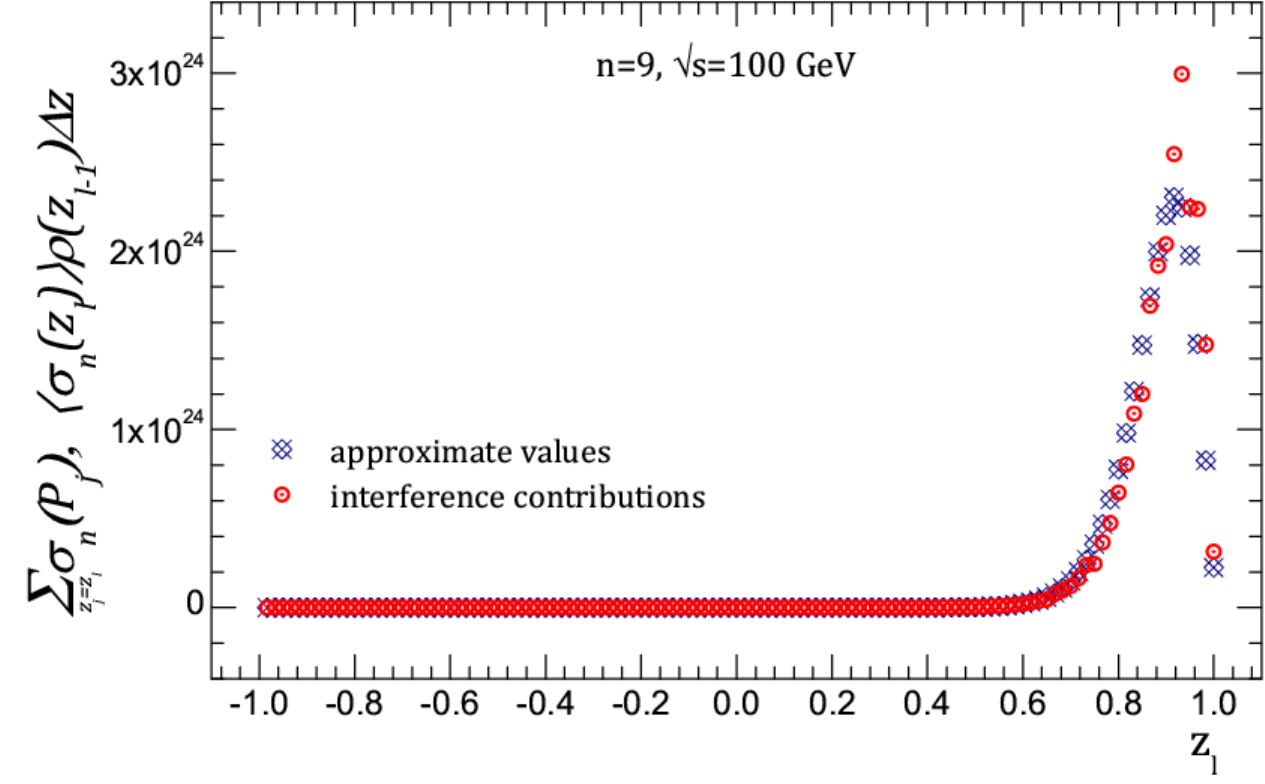} 
  \label{fig:part3_fig11d} 
  }
\end{center}
 \caption{
The values of $\sum\limits_{{z_j} = {z_l}} {{\sigma _n}\left( {{{\hat P}_j}} \right)}$ obtained with consideration of all interference contributions (circles) in comparision with the approximate values of $\left\langle {{\sigma _n}\left( {{z_l}} \right)} \right\rangle \rho \left( {{z_{l - 1}}} \right)\Delta z$ (crosses) for 
\ref{fig:part3_fig11a} - for $n = 8$ at  $\sqrt s = 10$ GeV,
\ref{fig:part3_fig11b} - for $n = 8$ at  $\sqrt s = 100$ GeV,
\ref{fig:part3_fig11c} - for $n = 9$ at  $\sqrt s = 10$ GeV,
\ref{fig:part3_fig11d} - for $n = 9$ at  $\sqrt s = 100$ GeV.}
  \label{fig:part3_fig11} 
\end{figure*}
%%%%%%%%%%%%---------------------------------%%%%%%%%%
%%%%%%%%%%%%---------------------------------%%%%%%%%%
%%%%%%%%%%%%---------------------------------%%%%%%%%%
%%%%%%%%%%%%---------------------------------%%%%%%%%%
\begin{figure*}
\begin{center}
  \centering
  \subfigure[]{
  \includegraphics[scale=0.79]{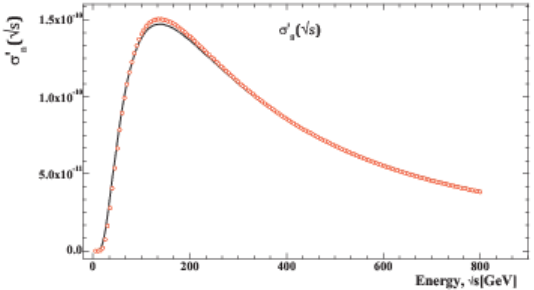} 
  \label{fig:part3_fig12a} 
  }
  \subfigure[]{
  \includegraphics[scale=0.79]{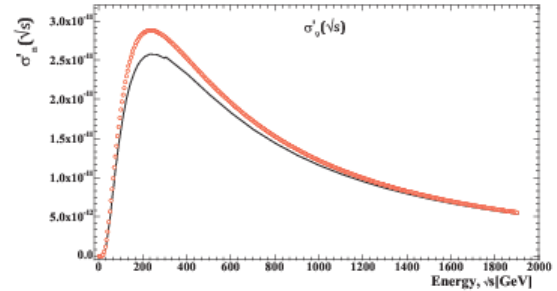} 
  \label{fig:part3_fig12b} 
  }
  \subfigure[]{
  \includegraphics[scale=0.79]{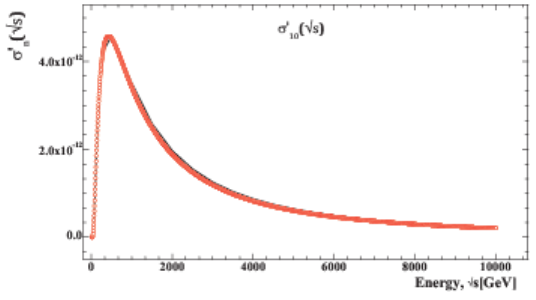} 
  \label{fig:part3_fig12c} 
  }
  \subfigure[]{
  \includegraphics[scale=0.74]{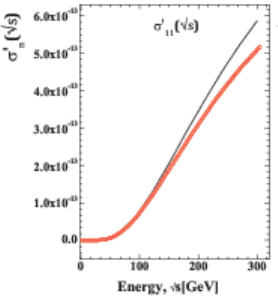} 
  \label{fig:part3_fig12d} 
  } 
  \subfigure[]{
  \includegraphics[scale=0.74]{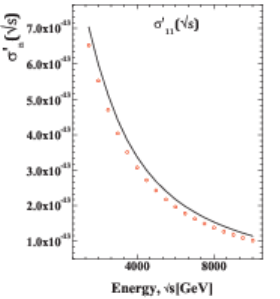} 
  \label{fig:part3_fig12e} 
  } 
\end{center}
 \caption{ 
The magnitude of
%The 
partial cross-section as a function of energy $\sqrt s$ calculated over all interference contributions (solid line) and by Eq.\ref{eq8} using approximations Eqs.\ref{eq17}, \ref{eq22}, \ref{eq23} (dashed line): \ref{fig:part3_fig12a} - $\sigma '_8( \sqrt s )$; \ref{fig:part3_fig12b} - $\sigma '_9( \sqrt s )$; \ref{fig:part3_fig12c} - $\sigma '_{10}( \sqrt s )$; \ref{fig:part3_fig12d} - $\sigma '_{11}( \sqrt s )$; \ref{fig:part3_fig12e} - $\sigma '_{11}( \sqrt s )$. Note, that this approximation is acceptable at least in the range of parameters in which they are can be verified.}
 \label{fig:part3_fig12}
\end{figure*}
%%%%%%%%%%%%---------------------------------%%%%%%%%%
%%%%%%%%%%%%---------------------------------%%%%%%%%%
%%%%%%%%%%%%---------------------------------%%%%%%%%%
%%%%%%%%%%%%---------------------------------%%%%%%%%%
\begin{figure*}
\begin{center}
  \centering
  \subfigure[]{
  \includegraphics[scale=0.37]{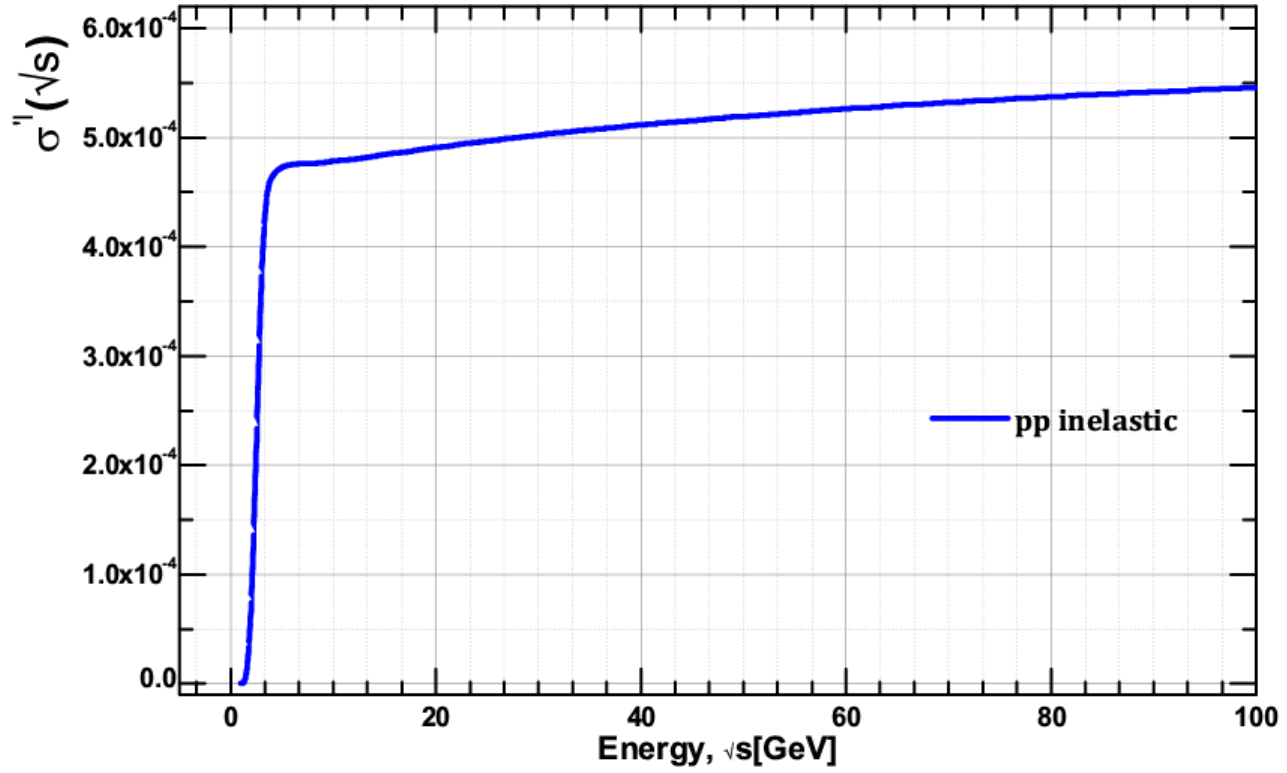} 
  \label{fig:part3_13a} 
  }
  \subfigure[]{
  \includegraphics[scale=0.35]{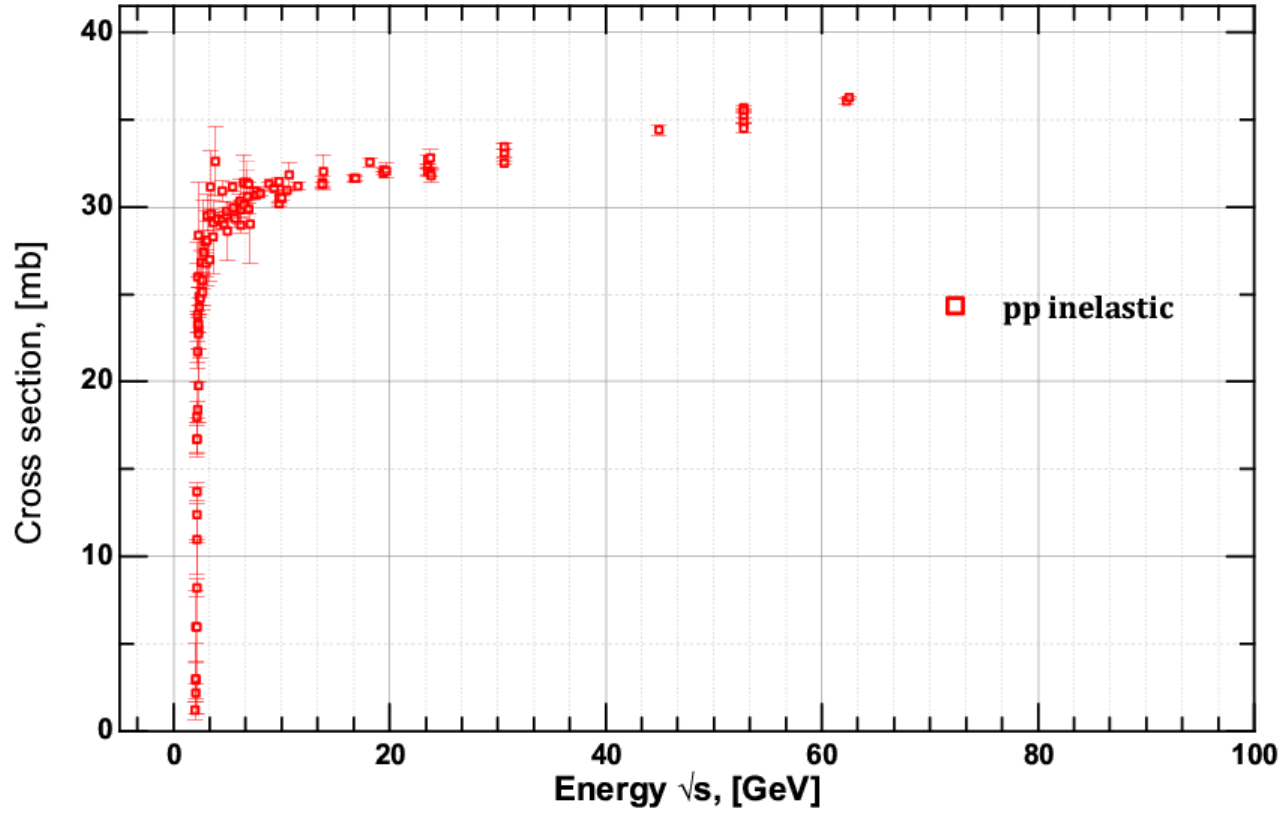} 
  \label{fig:part3_fig13c} 
  }
  \subfigure[]{
  \includegraphics[scale=0.37]{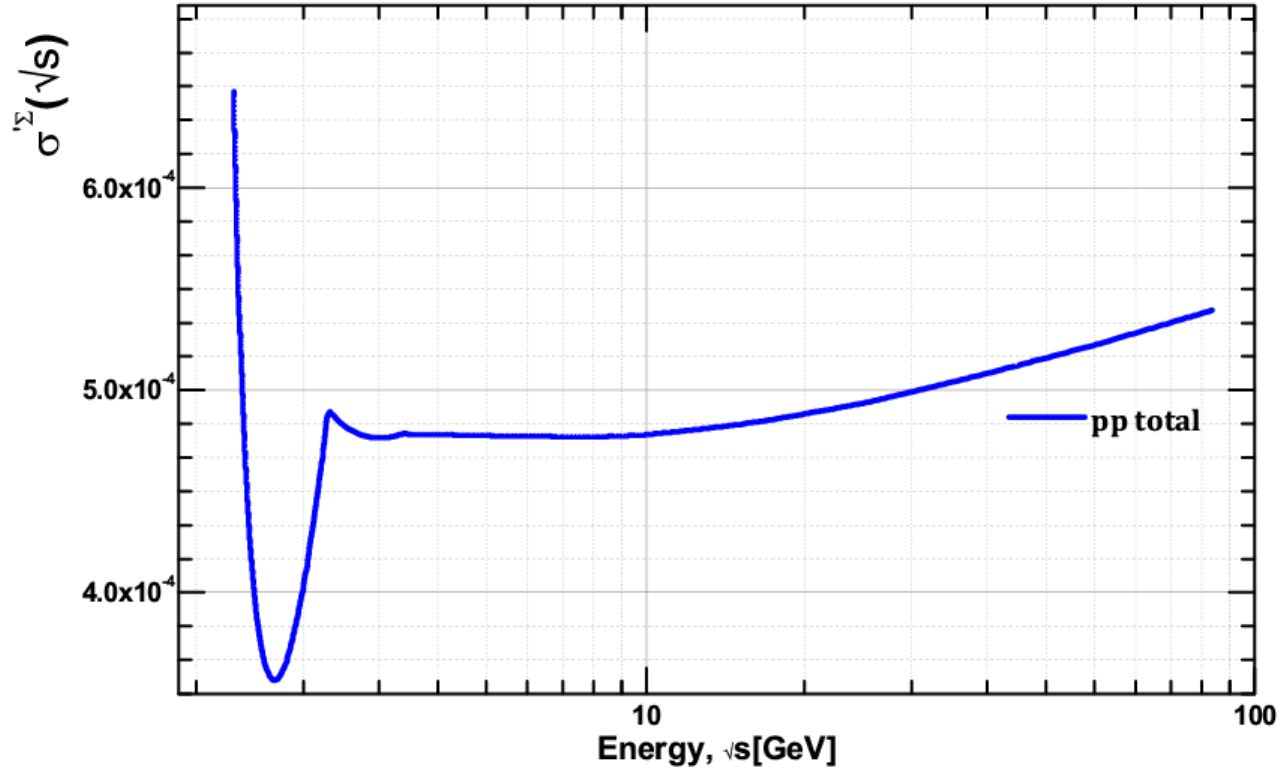} 
  \label{fig:part3_fig13b} 
  }
  \subfigure[]{
  \includegraphics[scale=0.35]{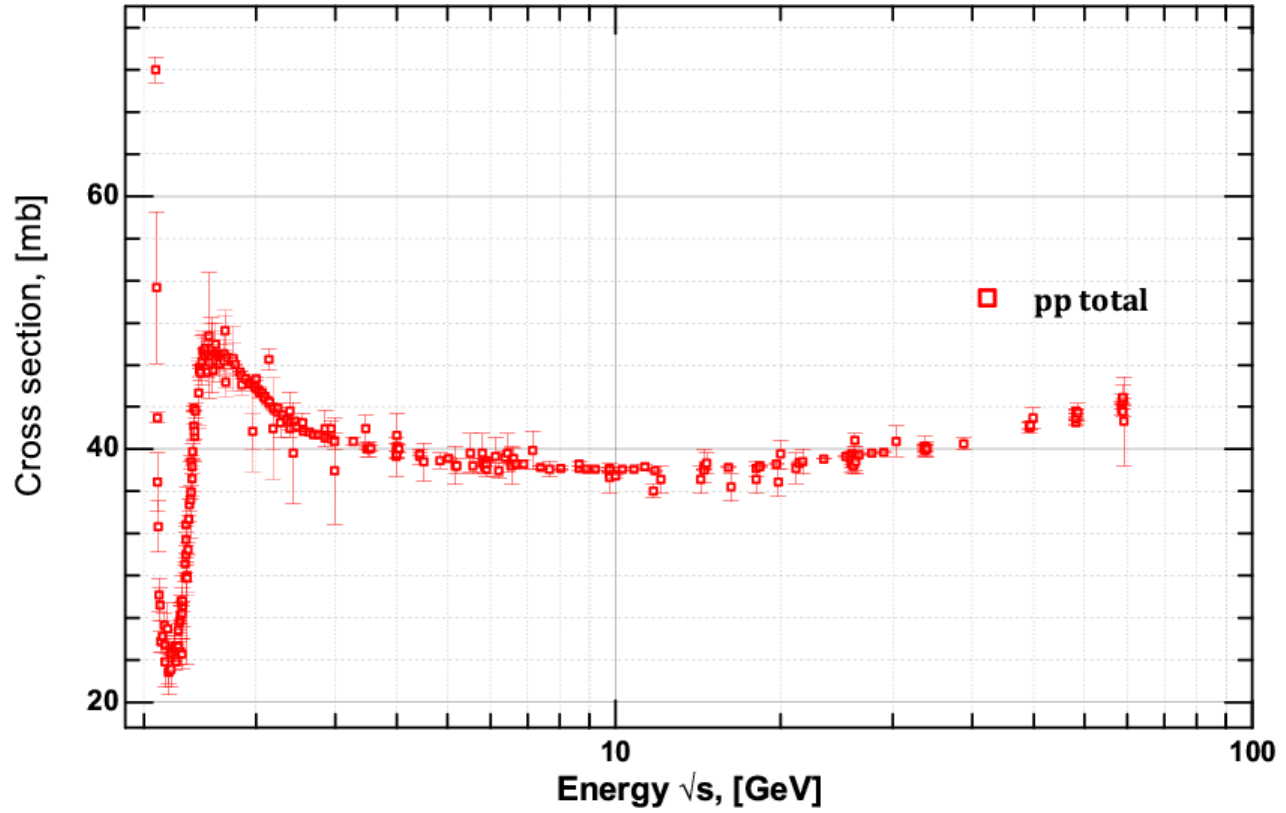} 
  \label{fig:part3_fig13d} 
  } 
\end{center}
 \caption{
Theoretical dependences of the $\sigma'^I(\sqrt s)$ \ref{fig:part3_13a} and $\sigma'^\Sigma(\sqrt s)$ \ref{fig:part3_fig13b} obtained for the energy range $\sqrt s =1\div100$ Gev at $L=5.51$. First minimum for the total cross-section can be obtained only when we take into account contributions from the high multiplicities. Experimental data for the inelastic \ref{fig:part3_fig13c} and for the total \ref{fig:part3_fig13d} pp scattering cross-section [\onlinecite{PDG_2010_JofPhysG, ATLASCollaboration_2011eu}] presented for qualitative comparison with the prediction from our model. Note: data-points for the inelastic cross-section, obtained from the definition $\sigma_{inel}=\sigma_{total}-\sigma_{elastic}$.}
 \label{fig:part3_fig13}
\end{figure*}
%%%%%%%%%%%%---------------------------------%%%%%%%%%
%%%%%%%%%%%%---------------------------------%%%%%%%%%
There is one more requirement, to which the areas obtained by partition of the sphere ${S_{n - 2}}$ must satisfy: they must not overlap, i.e., these regions do not have common internal points. Indeed, otherwise, at least any two of the examined arcs would intersect in some internal point of these arcs. As it follows from Eq.\ref{eq21}, when $n$ is large the value of ${r_n}$ is small. This means that when we join the each point of figure ${F_{n!}}$ with its nearest neighbors by the shortest arcs of sphere ${S_{n - 2}}$, these arcs practically coincide with chords, which tights them. 

If we assume, that any two chords ${A_{{i_1}}}{A_{{i_1} + 1}}$ and ${A_{{i_2}}}{A_{{i_2} + 1}}$ intersect in an internal point, then it is possible ``to pull'' on them a two-dimensional plane. Then we get a flat rectangle ${A_{{i_1}}}{A_{{i_2}}}{A_{{i_1} + 1}}{A_{{i_2} + 1}}$, which has at least one angle no smaller than $90^\circ $. This means that square of diagonal, lying opposite it is not less than sum of squares of the parties that make up the corner. Denoting the lengths of these sides through $a$ and $b$, we have ${a^2} + {b^2} \le r_n^2$. In this case, either $a$ or $b$ would not exceed ${r_n}/\sqrt 2 $, i.e., the figure ${F_{n!}}$ contains points, which are at distance less then ${r_n}$ but that cannot happen due to minimality of this distance.

Thus, we can conclude that at an arbitrary $n$ a sphere ${S_{n - 2}}$ can be partitioned into the parts of equal area, each of which contains only one point of figure ${F_{n!}}$, as it shown in Fig.\ref{fig:part3_fig07b}-\ref{fig:part3_fig07c}.

Let us introduce a multidimensional spherical coordinate system, so, that the end of vector ${\vec Y^{\left( 0 \right)}}$ is the ``north pole'' of sphere ${S_{n - 2}}$. Then the number of points of figure ${F_{n!}}$, to which the values of variable $z = \cos \left( \Theta  \right)$ in the interval $[z, z+dz]$ correspond, is equal 
%---------------------------------------------------------------------------------
\begin{eqnarray}
&& dN\left( {z,dz} \right) = \rho \left( z \right)dz 
\label{eq22}
\end{eqnarray}%
where
\begin{eqnarray}
&& \rho \left( z \right) = \frac{{n!}}{{\sqrt \pi  }}\frac{{\Gamma \left( {\frac{{n - 1}}{2}} \right)}}{{\Gamma \left( {\frac{{n - 2}}{2}} \right)}}{\left( {1 - {z^2}} \right)^{\frac{{n - 4}}{2}}} 
\end{eqnarray}%
%---------------------------------------------------------------------------------
$\Gamma $ is the Euler gamma function.

To verify the validity of Eq.\ref{eq22} we can calculate all interference contributions and corresponding values of $z$ at $n=8$ and $n=9$ (since for the larger number of particles this can not be realized). 

The distributions of interference contribution from the variable $z = \cos \left( \Theta  \right)$ and the graphs of function $\rho \left( z \right) = \frac{{dN\left( {z,z + dz} \right)}}{{dz}}$ from Eq.\ref{eq22} are shown in Fig.\ref{fig:part3_fig09}. Obtained results of numerical calculation of interference contributions and by Eq.\ref{eq22} are in a good agreement.

Moreover, as it follows from Fig.\ref{fig:part3_fig09b} and from Fig.\ref{fig:part3_fig09c} this fitness is improved with increasing number of particles $n$, i.e., Eq.\ref{eq22} is suitable for large $n$, when the direct numerical calculation of all interference contributions is impossible.

Taking Eq,\ref{eq22} and Eq.\ref{eq5} into account we obtain the following the approximate equality
%---------------------------------------------------------------------------------
\begin{eqnarray}
&& \Delta {N_l} \approx \rho \left( {{z_{l - 1}}} \right)\Delta z 
\label{eq23}
\end{eqnarray}%
%---------------------------------------------------------------------------------
where
%---------------------------------------------------------------------------------
\begin{eqnarray}
&& \Delta z = \frac{{12}}{{\left( {n - 1} \right)n\left( {n + 1} \right)}}
\end{eqnarray}%
%---------------------------------------------------------------------------------
Verification results of Eq.\ref{eq23} at $n=8$ and $n=9$ are presented in Fig.\ref{fig:part3_fig10}.

Another verification of considered above equations is presented in Fig.\ref{fig:part3_fig11}, where the values of $\sum\limits_{{z_j} = {z_l}} {{\sigma '_n}\left( {{{\hat P}_j}} \right)} $ and approximating magnitudes $\left\langle {{\sigma '_n}\left( {{z_l}} \right)} \right\rangle \rho \left( {{z_{l - 1}}} \right)\Delta z$  (here $\left\langle {{\sigma '_n}\left( {{z_l}} \right)} \right\rangle $ is calculated by Eq.\ref{eq17}) are compared.

From results demonstrated on Fig.\ref{fig:part3_fig04} and Fig.\ref{fig:part3_fig10}-\ref{fig:part3_fig12}, we can conclude that the at least for those numbers of particles for which it can be directly tested Eq.\ref{eq8} with Eq.\ref{eq17}, Eqs.\ref{eq22} - \ref{eq23} yields an acceptable approximation. As is obvious from Fig.\ref{fig:part3_fig04}, than closer energy to the threshold of $n$ particle production, the better approximation Eq.\ref{eq17}. Therefore, if we choose the range of low energies, for example, up to 100 GeV, because in this range total cross-section growth is observed, it is expected that the considered approximations will be acceptable for the large numbers of particles than those for which they were tested. In addition, as it follows from Fig.\ref{fig:part3_fig10b}-\ref{fig:part3_fig10d}, the accuracy of approximation Eq.\ref{eq23}, as expected, increases with the growth of $n$.

Thus, within the framework of examined approximations is possible to calculate the interference contributions at sufficiently large $n$, and we can consider the dependence of total inelastic cross-section on energy $\sqrt s $ in the simplest case of multi-peripheral model taking into account all significant interference contributions.

%#############################################
\section{The model of dependence of hadron inelastic scattering total cross-section on energy $\sqrt s$ }
\label{SECTION5}
%#############################################

Let us consider the magnitude
%---------------------------------------------------------------------------------
\begin{eqnarray}
&& {\sigma '^\Sigma }\left( {\sqrt s } \right) = \sum\limits_{n = 1}^{{n_{\max }}} {{L ^n}{\sigma' _n}\left( {\sqrt s } \right)} 
\label{eq24}
\end{eqnarray}%
%---------------------------------------------------------------------------------
which within the framework of the discussed above model is an analogue of total inelastic scattering cross-section. %, which is made nondimensional with the considered above magnitude const. 
Here ${n_{\max }}$ is the maximum number of secondary particles allowed by energy-momentum conservation law and $L$ is the dimensionless coupling constant, which we considered as a fitting parameter (see Eq.32 [\onlinecite{part2}]). Since the calculation of ${\sigma' _n}$ up to $n = {n_{\max }}$ takes a long time, so in practice we restrict the upper bound of summation by those values of $n$, beyond which the neglected contributions known to be smaller than the experimental error of cross-section measurements.

The constant $L$ can be fitted so that the dependence ${\sigma '_\Sigma }\left( {\sqrt s } \right)$ looks like the behavior of total hadron-hadron scattering cross-section with a minimum about $\sqrt s = 10$ GeV. The result of such a fitting is shown in Fig.\ref{fig:part3_fig13} (in that calculations we take proton mass as mass of primary particles and pion mass as mass of secondary particles).

Quantitative comparison with experimental data requires the consideration of more realistic model than the self-interacting scalar ${\phi ^3}$ field model.

%#############################################
\section{Conclusions}
\label{SECTION_Conclusion}
%#############################################

From obtained result, one might conclude that the considered in [\onlinecite{part1}] mechanism of virtuality reduction at the constrained maximum point of multi-peripheral scattering amplitude may be responsible for proton-proton total cross-section growth when all the considerable interference contributions are taken into account.

Just the revelation of mechanism of cross-section growth we consider as the main result of earlier papers [\onlinecite{part1, part2}] and present work, since this mechanism is intrinsic not only to the diagrams of the ``comb'' type, but also to different modifications of considered model.

Application the Laplace method allow to calculate another types of diagrams corresponding to various scenarios of hadron-hadron inelastic scattering and compare it with experimental data.

%###############################################
%\nocite{*}% dump all cites in Bib-file
%\section{}
%\label{REFERENCES}
\vspace{2cm} 
\normalsize {\textbf{REFERENCES}}
%\label{References}
%\bibliography{aipsamp}% Produces the bibliography via BibTeX.
\bibliography{References_JMPh}%% Produces the bibliography via BibTeX.
%###############################################

%#############################################
%\onecolumn
\appendix
%#############################################
%\section{Table 1}
\begin{table*}
\caption{
\label{fig:part3_table01}
Results of numerical calculations of the eigenvalues of matrix ${\hat D_y}$.}
\begin{ruledtabular}
\begin{tabular}{c|c|c|c|c|c}
%\hline 
\multicolumn{6}{c}{$n=20$} \\
\hline
\multicolumn{2}{c|}{$\sqrt s =10 $ GeV}& \multicolumn{2}{c|}{$\sqrt s =300 $ GeV}& \multicolumn{2}{c}{$\sqrt s =10 $ TeV}\\
\hline
$d_k^{\left( y \right)}$ & \mbox{\fontsize{10}{14}\selectfont $\frac{{d_k^{\left( y \right)} - \frac{1}{n}Sp\left( {{{\hat D}^{\left( y \right)}}} \right)}}{{\frac{1}{n}Sp\left( {{{\hat D}^{\left( y \right)}}} \right)}}$ } & $d_k^{\left( y \right)}$ & \mbox{\fontsize{10}{14}\selectfont $\frac{{d_k^{\left( y \right)} - \frac{1}{n}Sp\left( {{{\hat D}^{\left( y \right)}}} \right)}}{{\frac{1}{n}Sp\left( {{{\hat D}^{\left( y \right)}}} \right)}}$ } & $d_k^{\left( y \right)}$ & \mbox{\fontsize{10}{14}\selectfont $\frac{{d_k^{\left( y \right)} - \frac{1}{n}Sp\left( {{{\hat D}^{\left( y \right)}}} \right)}}{{\frac{1}{n}Sp\left( {{{\hat D}^{\left( y \right)}}} \right)}}$ } \\
\hline
1.317&	-0.417& 0.181& -0.864& 0.064& -0.928\\
3.078&	0.352& 0.551&	-0.586&	0.227&	-0.746\\
3.006&	0.321& 0.878&	-0.34&	0.421&	-0.527\\
1.883&	-0.173& 1.099&	-0.174&	0.604&	-0.321\\
2.53&	0.111& 1.238&	0.342&	0.745&	-0.163\\
2.527&	0.11& 1.785&	0.342&	0.849&	-0.047\\
2.401&	0.055& 1.785&	-0.07&	1.26&	0.415\\
2.399&	0.054& 1.324&	-0.005&	1.26&	0.415\\
2.061&	-0.094& 1.38&	0.037&	0.92&	0.033\\
2.312&	0.016& 1.416&	0.064&	0.967&	0.087\\
2.311&	0.015& 1.441&	0.083&	1.001&	0.124\\
2.124&	-0.067& 1.573&	0.183&	1.022&	0.147\\
2.248&	-0.012& 1.573&	0.183&	1.037&	0.164\\
2.247&	-0.013& 1.458&	0.096&	1.046&	0.175\\
2.152&	-0.055& 1.47&	0.105&	1.053&	0.183\\
2.161&	-0.05& 1.478&	0.111&	1.06&	0.19\\
2.203&	-0.032& 1.485&	0.116&	1.065&	0.196\\
2.203&	-0.032& 1.483&	0.115&	1.057&	0.188\\
2.174&	-0.045& 1.504&	0.131&	1.075&	0.207\\
\hline 
\multicolumn{6}{c}{$n=10$} \\
\hline
\multicolumn{2}{c|}{$\sqrt s =10 $ GeV}& \multicolumn{2}{c|}{$\sqrt s =300 $ GeV}& \multicolumn{2}{c}{$\sqrt s =10 $ TeV}\\
\hline
$d_k^{\left( y \right)}$ & \mbox{\fontsize{10}{14}\selectfont $\frac{{d_k^{\left( y \right)} - \frac{1}{n}Sp\left( {{{\hat D}^{\left( y \right)}}} \right)}}{{\frac{1}{n}Sp\left( {{{\hat D}^{\left( y \right)}}} \right)}}$ } & $d_k^{\left( y \right)}$ & \mbox{\fontsize{10}{14}\selectfont $\frac{{d_k^{\left( y \right)} - \frac{1}{n}Sp\left( {{{\hat D}^{\left( y \right)}}} \right)}}{{\frac{1}{n}Sp\left( {{{\hat D}^{\left( y \right)}}} \right)}}$ } & $d_k^{\left( y \right)}$ & \mbox{\fontsize{10}{14}\selectfont $\frac{{d_k^{\left( y \right)} - \frac{1}{n}Sp\left( {{{\hat D}^{\left( y \right)}}} \right)}}{{\frac{1}{n}Sp\left( {{{\hat D}^{\left( y \right)}}} \right)}}$ } \\
\hline
0.955&	-0.457&	0.147&	-0.809&	0.037&	-0.901\\
2.124&	0.207&	0.435&	-0.433&	0.13&	-0.65\\
2.121&	0.205&	0.665&	-0.133&	0.242&	-0.351\\
1.529&	-0.131&	0.794&	0.036&	0.34&	-0.087\\
1.707&	-0.03&	0.855&	0.115&	0.412&	0.107\\
1.77&	0.006&	0.882&	0.15&	0.46&	0.236\\
1.805&	0.026&	0.893&	0.164&	0.503&	0.352\\
1.891&	0.075&	1.052&	0.372&	0.489&	0.313\\
\end{tabular}
\end{ruledtabular}
\end{table*}

\end{document}